\documentclass[prd,preprint,tightenlines,floatfix,showpacs,preprintnumbers,nofootinbib,eqsecnum]{revtex4}

 \usepackage[dvips,final]{graphicx}
  \usepackage{amssymb}
   \usepackage{amsmath}
    \usepackage{amsfonts}
     \usepackage{epsfig}
      \usepackage{bm}

\begin{document}

\thispagestyle{empty} \preprint{\hbox{}} \vspace*{-10mm}

\title{Dijet correlations at RHIC,\\
leading-order $k_t$-factorization approach\\
versus next-to-leading order collinear approach}

\author{A.~Szczurek$^{1,2}$, A.~Rybarska$^{1}$ and G.~\'Slipek$^{1}$}

\email{antoni.szczurek@ifj.edu.pl}

\address{$^{1}$ {Institute of Nuclear Physics PAN, PL-31-342 Cracow, Poland}} 

\address{$^{2}$ {University of Rzesz\'ow, PL-35-959 Rzesz\'ow, Poland}}

\date{\today}

\begin{abstract}
We compare results of $k_t$-factorization approach and next-to-leading
order collinear-factorization approach for dijet correlations in
proton-proton collisions at RHIC energies.
We discuss correlations in azimuthal angle as well as 
correlations in two-dimensional space of transverse momenta of two jets.
Some $k_t$-factorization subprocesses are included for the first
time in the literature.
Different unintegrated gluon/parton distributions are used in
the $k_t$-factorization approach. The results depend on UGDF/UPDF used.
For collinear NLO case the situation depends significantly on whether we
consider correlations of any two jets or correlations of leading jets only.
In the first case the $2 \to 2$ contributions associated with soft
radiations summed up in the $k_t$-factorization approach dominate
at $\phi \sim \pi$ and at equal moduli of jet transverse momenta.
The collinear NLO $2 \to 3$ contributions dominate over
$k_t$-factorization cross section at small relative azimuthal
angles as well as for asymmetric transverse momentum configurations.
In the second case the NLO contributions vanish at small relative
azimuthal angles and/or large jet transverse-momentum disbalance due to
simple kinematical constraints.
There are no such limitations for the $k_t$-factorization approach.
All this makes the two approaches rather complementary.
The role of several cuts is discussed and quantified.
\end{abstract}


\maketitle

\section{Introduction}

The subject of jet correlations is interesting in the context
of recent detailed studies of hadron-hadron correlations
in nucleus-nucleus \cite{RHIC_correlations_nucleus_nucleus}
and proton-proton \cite{RHIC_correlations_proton_proton} collisions.
Those studies provide interesting information on the dynamics of
nuclear and elementary collisions.
Effects of geometrical jet structure were discussed recently
in Ref.\cite{Levai}. No QCD calculation of parton radiation was performed
up to now in this context. Before going into hadron-hadron correlations it
seems indispensable to understand better correlations between jets
due to the QCD radiation.
In this paper we address the case of elementary hadronic collisions in order
to avoid complicated and not yet well understood nuclear effects.
Our analysis should be considered as a first step in order to understand
the nuclear case in the future. We wish to address the problem how far
one can simplify the calculation to be useful and handy in the nuclear case
and yet realistic in the proton-proton case.

\begin{figure}[!h]    
 {\includegraphics[width=0.4\textwidth]{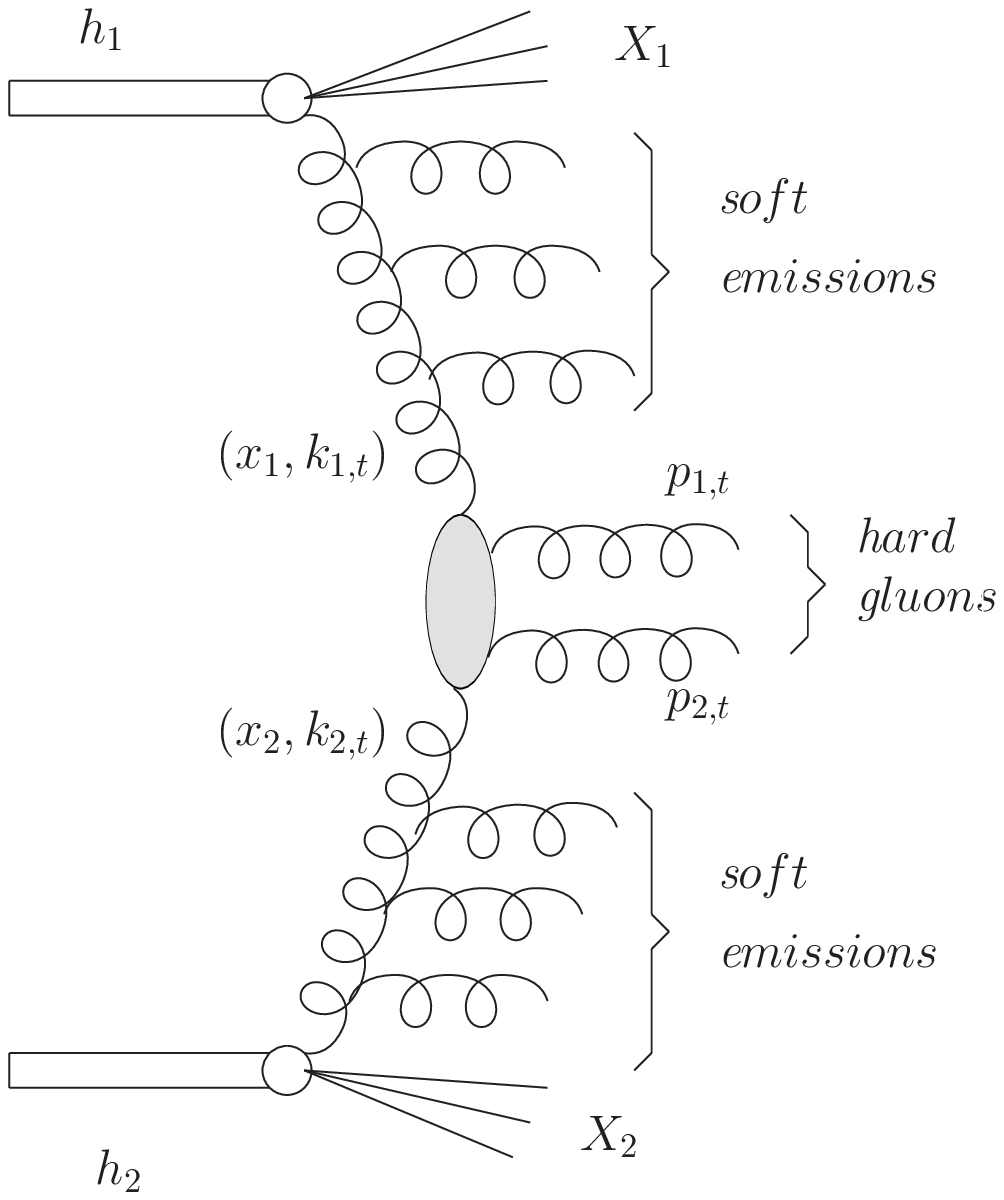}}
 {\includegraphics[width=0.4\textwidth]{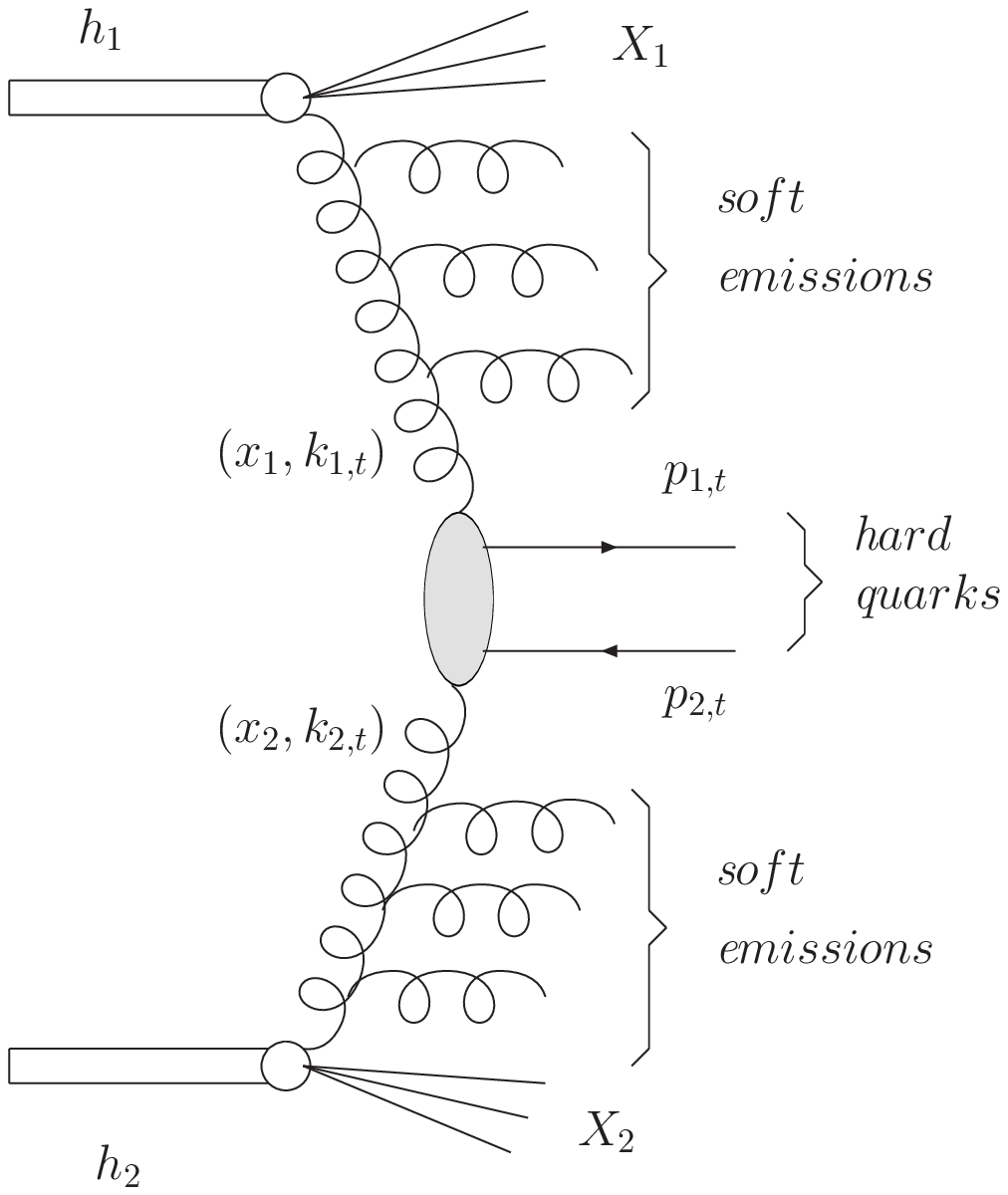}}
   \caption{\label{fig:kt_factorization_diagrams}
   \small  Typical diagrams for $k_t$-factorization approach to dijet
production.}
\end{figure}

In leading-order collinear-factorization approach jets are produced
back-to-back. These leading-order jets are therefore not included into
correlation function, although they contribute a big ($\sim
\frac{1}{2}$) fraction to the inclusive cross section.
The truly internal momentum distribution
of partons in hadrons due to Fermi motion (usually neglected in
the literature) and/or any soft emission would lead to a decorrelation
from the simple kinematical configuration.
In the fixed-order collinear approach only next-to-leading order terms
lead to nonvanishing cross sections at $\phi \ne \pi$ and/or
$p_{1,t} \ne p_{2,t}$ (moduli of transverse momenta of outgoing partons).
In the $k_t$-factorization approach, where transverse momenta
of gluons entering the hard process are included explicitly,
the decorrelations come naturally in a relatively easy to calculate way.
In Fig.\ref{fig:kt_factorization_diagrams} we show diagrams
illustrating the physics situation. The soft emissions, not explicit
in our calculation, are hidden in model unintegrated gluon distribution
functions (UGDF). In our calculation the last objects are assumed to be
given and are taken from the literature.


The $k_t$-factorization was originally proposed for heavy quark
production \cite{original_kt_factorization}.
In recent years it was used to describe several high-energy processes,
such as total cross section in virtual photon - proton scattering
\cite{UGDF_HERA}, heavy quark inclusive production\cite{BS00,LSZ02},
heavy quark -- heavy antiquark correlations \cite{LS04,LS06},
inclusive photon production \cite{LZ05_photon,PS07}, inclusive pion production
\cite{szczurek03,CS05}, Higgs boson \cite{Higgs} or gauge boson
\cite{KS04} production and dijet correlations in photoproduction
\cite{SNSS01} and hadroproduction \cite{LO00}.

It is often claimed that the $k_t$-factorization approach includes
implicitly some higher-order contributions of the standard collinear
approach. This loose statement requires a better understanding
and quantification.

Here we wish to address the problem of the relation between both approaches.
We shall identify the regions of the phase space where the
hard $2 \to 3$ processes, not explicitly included in the
leading-order $k_t$-factorization approach, dominate over the
$2 \to 2$ contributions calculated with UGDFs. 
We shall show how this depends on UGDFs used.  

We shall concentrate on the region of relatively semi-hard jets, i.e.
on the region related to the recently measured hadron-hadron
correlations at RHIC. Here the resummation effects may be expected
to be important. The resummation physics is addressed in our case
through the $k_t$-factorization approach.

\section{Formalism}

\subsection{$2 \to 2$ contributions with unintegrated
parton distributions}

It is known that at high energies, at midrapidities and not too large
transverse momenta the jet production is dominated by (sub)processes
initiated by gluons.
In this paper we concentrate only on such processes.
The region of forward/backward rapidities and/or processes with
large rapidity gap between jets will be studied elsewhere.
The cross section for the production of a pair of gluons or
a pair of quark-antiquark can be written as
\begin{eqnarray}
\frac{d\sigma(h_1 h_2 \rightarrow jj)}
{d^2p_{1,t}d^2p_{2,t}} &=& \int dy_1 dy_2
\frac{d^2 k_{1,t}}{\pi}\frac{d^2 k_{2,t}}{\pi}
\frac{1}{16\pi^2(x_1x_2s)^2}
\overline{|{\cal M}(gg\rightarrow jj)|^2}
\nonumber \\
&\cdot&\delta^2(\overrightarrow{k}_{1,t}
+\overrightarrow{k}_{2,t}
-\overrightarrow{p}_{1,t}
-\overrightarrow{p}_{2,t})
{\cal F}(x_1,k_{1,t}^2){\cal F}(x_2,k_{2,t}^2) \; ,
\label{basic_formula}
\end{eqnarray}
where
\begin{equation}
x_1 = \frac{m_{1,t}}{\sqrt{s}}\mathrm{e}^{+y_1} 
    + \frac{m_{2,t}}{\sqrt{s}}\mathrm{e}^{+y_2} \; ,
\end{equation}
\begin{equation}
x_2 = \frac{m_{1,t}}{\sqrt{s}}\mathrm{e}^{-y_1} 
    + \frac{m_{2,t}}{\sqrt{s}}\mathrm{e}^{-y_2} \; .
\end{equation}
The final partonic state is $jj = gg, q\bar q$.
If one makes the following replacement
\begin{equation}
{\cal F}_1(x_1,k_{1,t}^2) \rightarrow x_1g_1(x_1)
\delta(k_{1,t}^2)
\end{equation}
and
\begin{equation}
{\cal F}_2(x_2,k_{2,t}^2) \rightarrow x_2g_2(x_2)
\delta(k_{2,t}^2)
\end{equation}
then one recovers the familiar standard collinear-factorization formula.

The inclusive invariant cross section for $g$ production can be
written
\begin{equation}
\frac{d \sigma(h_1h_2 \to j)}{dy_1 d^2p_{1,t}} = 2
\int dy_2 \frac{d^2 k_{1,t}}{\pi} \frac{d^2 k_{2,t}}{\pi}
\left( ... \right) |
_{\vec{p}_{2,t} = \vec{k}_{1,t} + \vec{k}_{2,t} - \vec{p}_{1,t}}
\label{inclusive1}
\end{equation}
and equivalently as
\begin{equation}
\frac{d \sigma(h_1h_2 \to j)}{dy_2 d^2p_{2,t}} = 2
\int dy_1 \frac{d^2 k_{1,t}}{\pi} \frac{d^2 k_{2,t}}{\pi}
\left( ... \right) |
_{\vec{p}_{1,t} = \vec{k}_{1,t} + \vec{k}_{2,t} -
  \vec{p}_{2,t}} \;.
\label{inclusive2}
\end{equation}

Let us return to the coincidence cross section.
The integration with the Dirac delta function in (\ref{basic_formula})
\begin{equation}
\int dy_1 dy_2
\frac{d^2 k_{1,t}}{\pi}\frac{d^2 k_{2,t}}{\pi}
\left(...\right) \delta^2(...)\; .
\end{equation}
can be performed by introducing the following new auxiliary variables:
\begin{eqnarray}
\overrightarrow{Q}_t &=&
\overrightarrow{k}_{1,t}+\overrightarrow{k}_{2,t} \; ,
\nonumber \\
\overrightarrow{q}_t &=&
\overrightarrow{k}_{1,t}-\overrightarrow{k}_{2,t} \; .
\end{eqnarray}
The jacobian of this transformation is:
\begin{equation}
\frac{\partial 
(\overrightarrow{Q}_t,\overrightarrow{q}_t)}
{\partial
(\overrightarrow{k}_{1,t},
\overrightarrow{k}_{2,t})}=
\begin{pmatrix}
1 & 1 \\
1 &-1 \\
\end{pmatrix}
\cdot
\begin{pmatrix}
1 & 1 \\
1 &-1 \\
\end{pmatrix}
=2 \cdot 2 = 4 \; .
\end{equation}
Then our initial cross section can be written as:
\begin{equation}
\frac{d\sigma(h_1 h_2 \rightarrow Q \bar Q)}
{d^2p_{1,t}d^2p_{2,t}} =
\frac{1}{4} \int dy_1 dy_2
\;d^2Q_t d^2q_t \; ( ... ) \;
\delta^2(\overrightarrow{Q}_t-\overrightarrow{p}_{1,t}
-\overrightarrow{p}_{2,t})
\end{equation}
\begin{equation}
= \frac{1}{4} \int dy_1 dy_2\; 
\underbrace{d^2q_t}\; \left(...\right) \;
|_{\overrightarrow{Q}_t=\overrightarrow{P}_t} = 
\end{equation}
\begin{equation}
= \frac{1}{4} \int dy_1 dy_2\; 
\overbrace{\underbrace{q_tdq_t} \;d\phi_{q_t}} \; \left(...\right) \; 
|_{\overrightarrow{Q}_t=\overrightarrow{P}_t} =
\end{equation}
\begin{equation}
= \frac{1}{4} \int dy_1 dy_2\;
\overbrace{\frac{1}{2}dq_t^2 \;d\phi_{q_t}} \; \left(...\right) \;
|_{\overrightarrow{Q}_t=\overrightarrow{P}_t} \;.
\end{equation}
Above $\vec{P}_t = \vec{p}_{1,t} + \vec{p}_{2,t}$.
Different representations of the cross section are
possible. If one is interested in the distribution of the sum of
transverse momenta of the outgoing quarks, then it is convenient
to write 
\begin{eqnarray}
d^2p_{1,t}\; d^2p_{2,t} &=& \frac{1}{4}d^2P_td^2p_t = 
\frac{1}{4}d\phi_{P_t}P_tdP_t \;d\phi_{p_t}p_tdp_t
\nonumber \\
&=&\frac{1}{4}\;2\pi P_tdP_t \;d\phi_{p_t}p_tdp_t \; .
\end{eqnarray}
If one is interested in studying a two-dimensional map
$p_{1,t} \times p_{2,t}$ then
\begin{equation}
d^2p_{1,t}\; d^2p_{2,t} = 
d\phi_1 \; p_{1,t} dp_{1,t} \; 
d\phi_2 \; p_{2,t} dp_{2,t} \; .
\end{equation}
Then the two-dimensional map in jets transverse momenta can be written as
\begin{equation}
\frac{d \sigma(p_{1,t},p_{2,t})}{d p_{1,t} d p_{2,t}}
= \int d \phi_1 d \phi_2 \; p_{1,t} p_{2,t} \;
\int d y_1 d y_2 \; \frac{1}{4} q_t d q_t d \phi_{q_t} \left(...\right) \; .
\label{2-dim-map_a}
\end{equation}
The integral over $\phi_1$ and $\phi_2$ must be the most external one.
The integral above is formally a 6-dimensional one.
It is convenient to make the following transformation of variables
\begin{equation}
(\phi_1, \phi_2) \to \left( \phi_{+} = \phi_1 + \phi_2, 
\; \phi_{-} = \phi_1 - \phi_2 \right) \; ,
\end{equation}
where $\phi_{+} \in (0,4\pi)$ and $\phi_{-} \in (-2\pi,2\pi)$.
Now the new domain $(\phi_{+},\phi_{-})$
is twice bigger than the original one $(\phi_1,\phi_2)$.
The differential element
\begin{equation}
d\phi_1d\phi_2 = \left(
\frac{\partial \phi_1 \partial \phi_2}
  {\partial \phi_{+} \partial \phi_{-}} \right)
d \phi_{+} d \phi_{-}  \; .
\end{equation}
The transformation jacobian is:
\begin{equation}
\left(
\frac
{\partial \phi_1 \partial \phi_2}
{\partial \phi_{+} \partial \phi_{-}}
\right) 
= \frac{1}{2} \; .
\label{jacobian_in_phis}
\end{equation}
Then
\begin{eqnarray}
d^2p_{1,t}\; d^2p_{2,t} &=& 
=p_{1,t}dp_{1,t}\;p_{2,t}dp_{2,t}
{\frac{d\phi_{+}d\phi_{-}}{2}}
\nonumber \\
&=& p_{1,t}dp_{1,t}\;p_{2,t}dp_{2,t} \; 2\pi d\phi_{-} \; .
\end{eqnarray}

The integrals in Eq.(\ref{2-dim-map_a}) can be written equivalently as
\begin{equation}
\frac{d \sigma(p_{1,t},p_{2,t})}{d p_{1,t} d p_{2,t}}
= \frac{1}{2} \cdot \frac{1}{2}
\int d \phi_{+} d \phi_{-} \; p_{1,t} p_{2,t} \;
\int d y_1 d y_2 \; \frac{1}{4} q_t d q_t d \phi_{q_t} \left(...\right) \; .
\label{2-dim-map_b}
\end{equation}
The first factor of $\frac{1}{2}$ comes from the jacobian of
the transformation and the second $\frac{1}{2}$ is due to the extra
extension of the domain.

By symmetry, there is no dependence on $\phi_{+}$ and therefore
the final result can be written as:
\begin{equation}
\frac{d \sigma(p_{1,t},p_{2,t})}{d p_{1,t} d p_{2,t}}
= \frac{1}{2} \cdot \frac{1}{2} \cdot 4 \pi
\int d \phi_{-} \; p_{1,t} p_{2,t} \;
\int d y_1 d y_2 \; \frac{1}{4} q_t d q_t d \phi_{q_t} \left(...\right) \; .
\label{2-dim-map_c}
\end{equation}
This 5-dimensional integral is now calculated for each point on the map
$p_{1,t} \times p_{2,t}$.
This formula can be also used to calculate a single particle spectrum
of parton 1 and parton 2.

The matrix elements for 2 $\to$ 2 processes are discussed shortly
in Appendix A. The analytical continuation of the
standard on-shell matrix elements (see Appendix A) will be called in
the following ``on-shell approximation'' for brevity.
In Refs.\cite{LO00,O00} exact matrix elements for off-shell 
initial gluons were presented (see Appendix A).
We have checked that the results obtained with the on-shell
approximation and those obtained with the off-shell matrix elements
are numerically almost identical. The deviations occur only for very
virtual (large $k_t$) gluons where the contribution to
the cross section is small for majority of UGDFs.

\begin{figure}[!h]    
 {\includegraphics[width=0.4\textwidth]{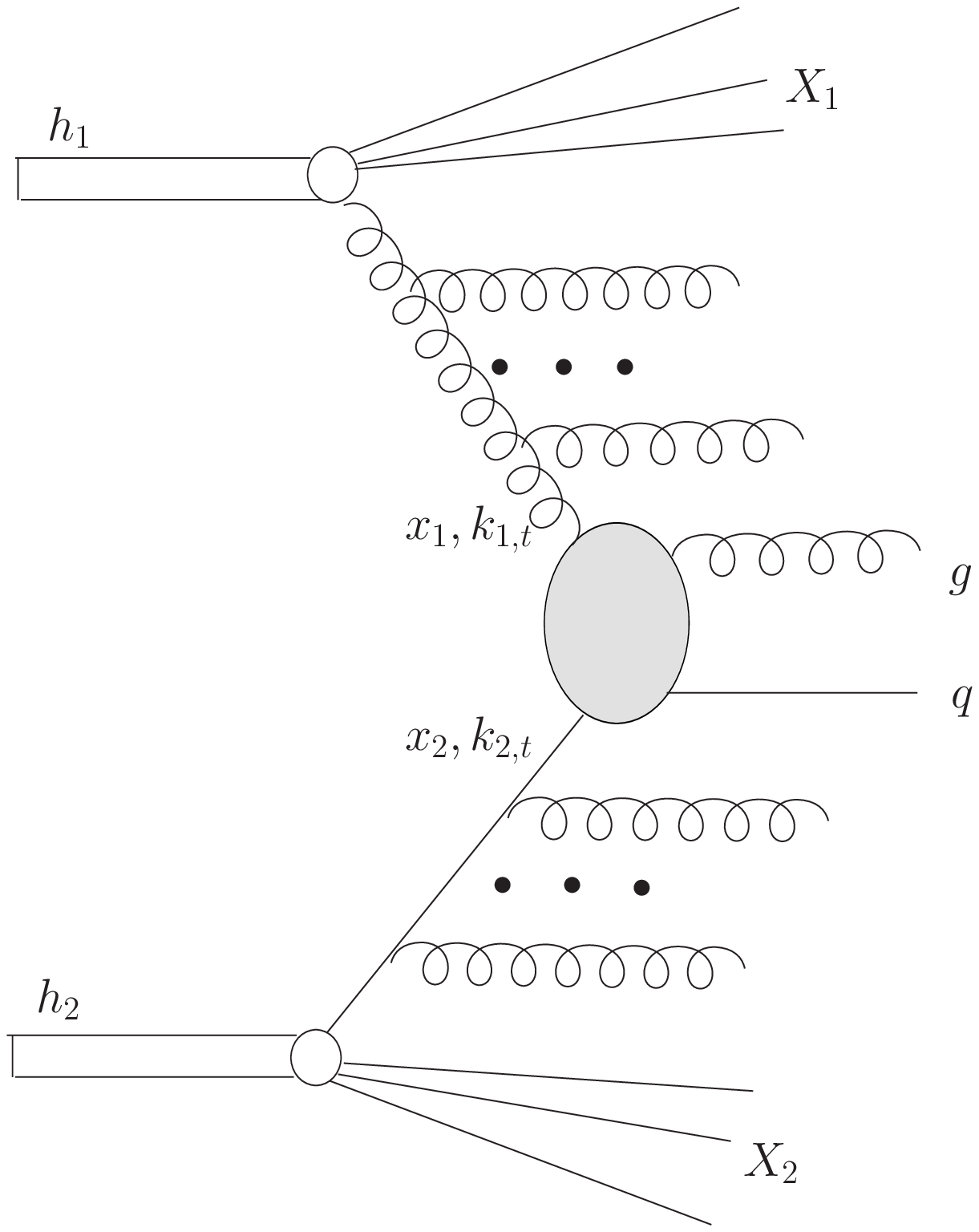}}
 {\includegraphics[width=0.4\textwidth]{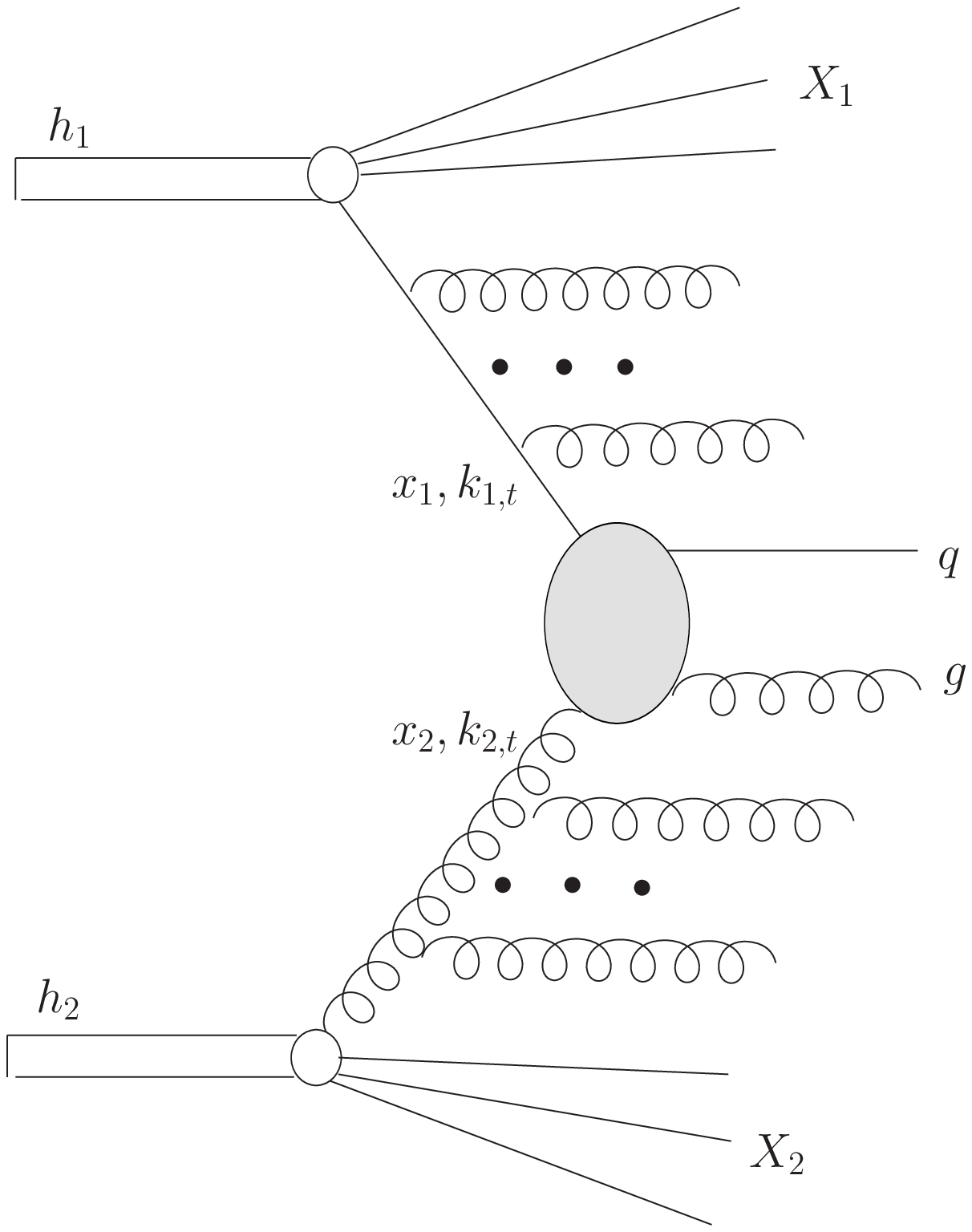}}
   \caption{\label{fig:new_contributions}
   \small  New $k_t$-factorization contributions included in the present
paper}
\end{figure}

In the present calculation we shall include also components with
gluon-quark and quark-gluon processes shown in
Fig.\ref{fig:new_contributions}. In the next section we shall discuss
how large are their contributions to the cross section.

\subsection{$2 \to 3$ contributions in collinear-factorization
approach}

\begin{figure}[!h]    
 \centerline{\includegraphics[width=0.8\textwidth]{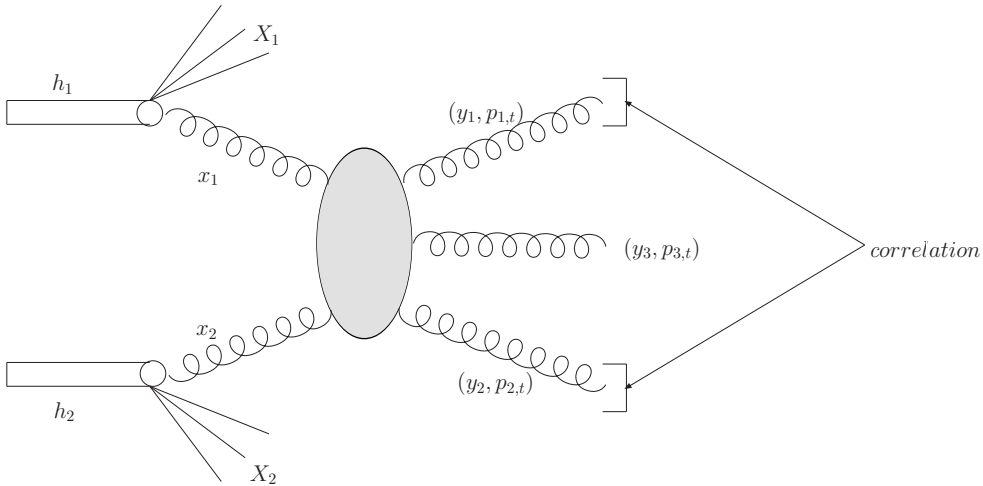}}
   \caption{\label{fig:NLO_diagrams}
   \small  A typical diagram for $2 \to 3$ contributions.
The kinematical variables used are shown explicitly.}
\end{figure}

Up to now we have considered only processes with two explicit hard
partons. In this section we shall discuss processes with three explicit
hard partons.
In Fig.\ref{fig:NLO_diagrams} we show a typical $2 \to 3$ process.
We also show kinematical variables needed in the description of
the process. We select the particle 1 and 2 as those which correlations
are studied. This is only formal as all possible combinations are
considered in real calculations.


The cross section for $h_1 h_2 \to g g g X$ can be calculated according
to the standard parton model formula:
\begin{equation}
d \sigma (h_1 h_2 \to g g g) =
\int d x_1 d x_2 \; g_1(x_1,\mu^2) g_2(x_2,\mu^2) \; 
d {\hat \sigma}(g g \to g g g)
\label{2to3_parton_formula}
\end{equation}
The elementary cross section can be written as
\begin{equation}
d {\hat \sigma}(g g \to g g g) = \frac{1}{2 {\hat s}}
\overline{| {\cal M}_{gg \to ggg} |^2} d R_3 \; .
\label{2to3_elementary_cross_section}
\end{equation}
The three-body phase space element is:
\begin{equation}
d R_3 =
\frac{ d^3 p_1 }{2E_1 (2 \pi)^3}
\frac{ d^3 p_2 }{2E_2 (2 \pi)^3}
\frac{ d^3 p_3 }{2E_3 (2 \pi)^3}
(2 \pi)^4 \delta^4(p_a+p_b-p_1-p_2-p_3) \; ,
\end{equation}
It can be written in an equivalent way in terms of parton rapidities 
\begin{equation}
d R_3 =
\frac{ d y_1 d^2 p_{1,t} }{(4 \pi) (2 \pi)^2}
\frac{ d y_2 d^2 p_{2,t} }{(4 \pi) (2 \pi)^2}
\frac{ d y_3 d^2 p_{3,t} }{(4 \pi) (2 \pi)^2}
(2 \pi)^4 \delta^4(p_a+p_b-p_1-p_2-p_3) \; .
\end{equation}
The last formula is useful for practical purposes.
Now the cross section for hadronic collisions can be written
in terms of $2 \to 3$ matrix element as
\begin{equation}
d \sigma = d y_1 d^2 p_{1,t} d y_2 d^2 p_{2,t} d y_3
\cdot \frac{1}{(4 \pi)^3 (2 \pi)^2} \; \frac{1}{\hat{s}^2} \;
x_1 f_1(x_1,\mu_f^2) x_2 f_2(x_2,\mu_f^2) \; 
\overline{|{\cal M}_{2 \to 3}|^2} \; ,
\end{equation}
where the longitudinal momentum fractions are evaluated as
\begin{eqnarray}
x_1 &=& \frac{p_{1,t}}{\sqrt{s}} \exp(+y_1)
      + \frac{p_{2,t}}{\sqrt{s}} \exp(+y_2)
      + \frac{p_{3,t}}{\sqrt{s}} \exp(+y_3) \; ,\nonumber \\
x_2 &=& \frac{p_{1,t}}{\sqrt{s}} \exp(-y_1)
      + \frac{p_{2,t}}{\sqrt{s}} \exp(-y_2)
      + \frac{p_{3,t}}{\sqrt{s}} \exp(-y_3) \; .
\end{eqnarray}
Repeating similar steps as for $2 \to 2$ processes we get finally:
\begin{equation}
d \sigma = \frac{1}{64 \pi^4 \hat{s}^2} \; 
x_1 f_1(x_1,\mu_f^2) x_2 f_2(x_2,\mu_f^2) \; \overline{|{\cal M}_{2 \to 3}|^2}
p_{1,t} dp_{1,t} p_{2,t} dp_{2,t} d \phi_{-} dy_1 dy_2 dy_3 \; ,
\label{2to3_handy_formula}
\end{equation}
where $\phi_{-}$ is restricted to the interval $(0,\pi)$.
The last formula is very useful in calculating the cross section
for particle 1 and particle 2 correlations.

\subsection{Unintegrated gluon distributions}

In general, there are no simple relations between unintegrated
and integrated parton distributions.
Some of UPDFs in the literature are obtained based on familiar
collinear distributions, some are obtained by solving evolution
equations, some are just modelled or some are even parametrized.
A brief review of unintegrated gluon distributions (UGDFs) that will
be used here can be found in Ref.\cite{LS06}.
We shall not repeat all details concerning those UGDFs here.
We shall discuss in more details only approaches which treat
unintegrated quark/antiquark distributions.

In some of approaches one imposes the following relation
between the standard collinear distributions and UPDFs:
\begin{equation}
a(x,\mu^2) = \int_0^{\mu^2} f_a(x,{\mathrm {\bf k}}_t^2,\mu^2)
\frac{d{\mathrm {\bf k}}_t^2}{{\mathrm {\bf k}}_t^2}   \; ,
\end{equation}
where $a = xq$ or $a = xg$.

Since familiar collinear distributions satisfy sum rules, one can
define and test analogous sum rules for UPDFs.
We shall discuss this issue in more detail in a separate section. 

Due to its simplicity the Gaussian smearing of initial transverse momenta
is a good reference for other approaches. It allows to study
phenomenologically the role of transverse momenta in several
high-energy processes.
We define a simple unintegrated parton distributions:
\begin{equation}
{\cal F}_{i}^{Gauss}(x,k^2,\mu_F^2) = x p_{i}^{coll}(x,\mu_F^2)
\cdot f_{Gauss}(k^2) \; ,
\label{Gaussian_UPDFs}
\end{equation}
where $p_{i}^{coll}(x,\mu_F^2)$ are standard collinear (integrated)
parton distribution ($i = g, q, \bar q$) and $f_{Gauss}(k^2)$
is a Gaussian two-dimensional function:
\begin{equation}
f_{Gauss}(k^2) = \frac{1}{2 \pi \sigma_0^2}
\exp \left( -k_t^2 / 2 \sigma_0^2 \right)  \frac{1}{\pi} \; .
\label{Gaussian}
\end{equation}
The UPDFs defined by Eq.(\ref{Gaussian_UPDFs}) and (\ref{Gaussian})
is normalized such that:
\begin{equation}
\int {\cal F}_{i}^{Gauss}(x,k^2,\mu_F^2) \; d k^2 = x
p_{i}^{coll}(x,\mu_F^2) \; .
\label{Gaussian_normalization}
\end{equation}

Kwieci\'nski has shown that the evolution equations
for unintegrated parton distributions takes a particularly
simple form in the variable conjugated to the parton transverse momentum.
In the impact-parameter space the Kwieci\'nski equations
takes the following relatively simple form
\begin{equation}
\begin{split}
{\partial{\tilde {\cal F}_{NS}(x,b,\mu^2)}\over \partial \mu^2} &=
{\alpha_s(\mu^2)\over 2\pi \mu^2}  \int_0^1dz  \, P_{qq}(z)
\bigg[\Theta(z-x)\,J_0((1-z) \mu b)\,
{\tilde {\cal F}_{NS}\left({x\over z},b,\mu^2 \right)}
\\&- {\tilde {\cal F}_{NS}(x,b,\mu^2)} \bigg]  \; , \\
{\partial{\tilde {\cal F}_{S}(x,b,\mu^2)}\over \partial \mu^2} &=
{\alpha_s(\mu^2)\over 2\pi \mu^2} \int_0^1 dz
\bigg\{\Theta(z-x)\,J_0((1-z) \mu b)\bigg[P_{qq}(z)\,
 {\tilde {\cal F}_{S}\left({x\over z},b,\mu^2 \right)}
\\&+ P_{qg}(z)\, {\tilde {\cal F}_{G}\left({x\over z},b,\mu^2 \right)}\bigg]
 - [zP_{qq}(z)+zP_{gq}(z)]\,
{\tilde {\cal F}_{S}(x,b,\mu^2)}\bigg\}  \; ,
 \\
{ \partial {\tilde {\cal F}_{G}(x,b,\mu^2)}\over \partial \mu^2}&=
{\alpha_s(\mu^2)\over 2\pi \mu^2} \int_0^1 dz
\bigg\{\Theta(z-x)\,J_0((1-z) \mu b)\bigg[P_{gq}(z)\,
{\tilde {\cal F}_{S}\left({x\over z},b,\mu^2 \right)}
\\&+ P_{gg}(z)\, {\tilde {\cal F}_{G}\left({x\over z},b,\mu^2 \right)}\bigg]
-[zP_{gg}(z)+zP_{qg}(z)]\, {\tilde {\cal F}_{G}(x,b,\mu^2)}\bigg\} \; .
\end{split}
\label{kwiecinski_equations}
\end{equation}
We have introduced here the short-hand notation
\begin{equation}
\begin{split}
\tilde {\cal F}_{NS}&= \tilde {\cal F}_u - \tilde {\cal F}_{\bar u}, \;\;
                 \tilde {\cal F}_d - \tilde {\cal F}_{\bar d} \; ,  \\
\tilde {\cal F}_{S}&= \tilde {\cal F}_u + \tilde {\cal F}_{\bar u} + 
                \tilde {\cal F}_d + \tilde {\cal F}_{\bar d} + 
                \tilde {\cal F}_s + \tilde {\cal F}_{\bar s} \; . 
\end{split}
\label{singlet_nonsinglet}
\end{equation}
The unintegrated parton distributions in the impact factor
representation are related to the familiar collinear distributions
as follows
\begin{equation}
\tilde {\cal F}_{k}(x,b=0,\mu^2)=\frac{x}{2} p_k(x,\mu^2) \; .
\label{uPDF_coll_1}
\end{equation}
On the other hand, the transverse momentum dependent UPDFs are related
to the integrated parton distributions as
\begin{equation}
x p_k(x,\mu^2) =
\int_0^{\infty} d k_t^2 \; {\cal F}_k(x,k_t^2,\mu^2) \; .
\label{uPDF_coll_2}
\end{equation}

The two possible representations are interrelated via Fourier-Bessel
transform
\begin{equation}
  \begin{split}
    &{{\cal F}_k(x,k_t^2,\mu^2)} =
    \int_{0}^{\infty} db \;  b J_0(k_t b)
    {{\tilde {\cal F}}_k(x,b,\mu^2)} \; ,
    \\
    &{{\tilde {\cal F}}_k(x,b,\mu^2)} =
    \int_{0}^{\infty} d k_t \;  k_t J_0(k_t b)
    {{\cal F}_k(x,k_t^2,\mu^2)} \; .
  \end{split}
\label{Fourier}
\end{equation}
The index k above numerates either gluons (k=0), quarks (k$>$ 0) or
antiquarks (k$<$ 0).

While physically ${\cal F}_k(x,k_t^2,\mu^2)$ should be positive,
there is no obvious reason for such a limitation for
$\tilde {\cal F}_k(x,b,\mu^2)$.

In the following we use leading-order parton distributions
from Ref.\cite{GRV98} as the initial condition for QCD evolution.
The set of integro-differential equations in b-space
was solved by the method based on the discretisation made with
the help of the Chebyshev polynomials (see \cite{kwiecinski}).
Then the unintegrated parton distributions were put on a grid
in $x$, $b$ and $\mu^2$ and the grid was used in practical
applications for Chebyshev interpolation. 

For the calculation of jet correlations here the parton distributions
in momentum space are more useful.
These calculation requires a time-consuming multi-dimensional
integration. An explicit calculation of the Kwieci\'nski UPDFs
via Fourier transform for needed in the main calculation values of
$(x_1,k_{1,t}^2)$ and $(x_2,k_{2,t}^2)$ (see next section)
is not possible.
Therefore auxiliary grids of
the momentum-representation UPDFs are prepared before
the actual calculation of the cross sections.
These grids are then used via a two-dimensional interpolation
in the spaces $(x_1,k_{1,t}^2)$ and $(x_2,k_{2,t}^2)$
associated with each of the two incoming partons.

%
%

\section{Results}

Let us concentrate first on $2 \to 2$ processes calculated with
the inclusion of initial transverse momenta.
We shall include the following four (sub)processes:
\begin{itemize}
\item{gluon+gluon $\to$ gluon+gluon} (called diagram $A_1$, see Fig.1a)
\item{gluon+gluon $\to$ quark+antiquark} (called diagram $A_2$, see Fig.1b)
\item{gluon+(anti)quark $\to$ gluon+(anti)quark} (called diagram $B_1$, 
see Fig.2a)
\item{(anti)quark+gluon $\to$ (anti)quark+gluon} (called diagram $B_2$, 
see Fig.2b)
\end{itemize}
Only first two were included recently in the $k_t$-factorization
approach \cite{LO00,Bartels}.
The papers in the literature have been concentrated on large energies, i.e.
on such cases when only gluons come into game. We shall show that
at present subasymptotic energies (RHIC, Tevatron) also the last two
must be included, even at midrapidities. Similar conclusion was drawn
recently for inclusive pion distributions at RHIC \cite{CS05}.

\begin{figure}[!hp]      
\begin{minipage}{0.49\textwidth}
\epsfxsize=\textwidth\epsfbox{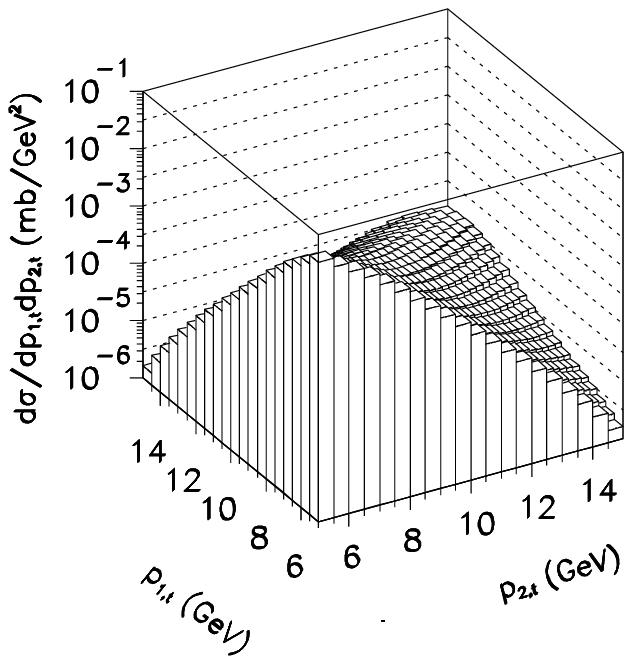}
\end{minipage}
\begin{minipage}{0.49\textwidth}
\epsfxsize=\textwidth \epsfbox{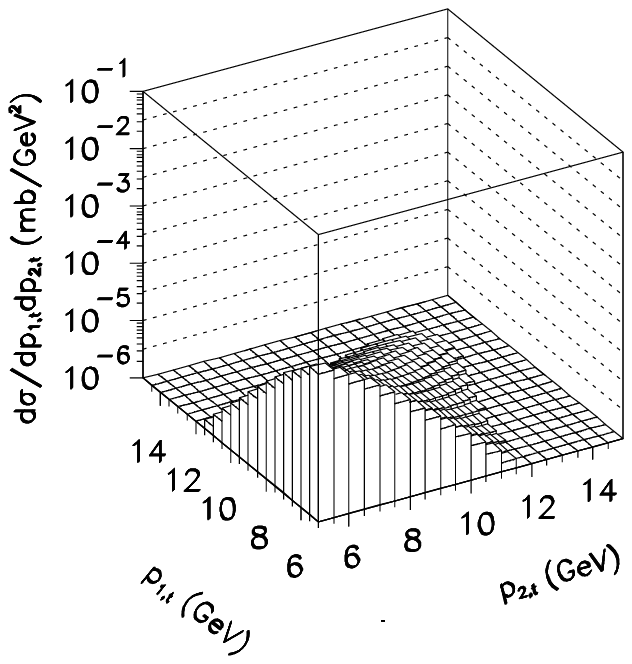}
\end{minipage}
\begin{minipage}{0.49\textwidth}
\epsfxsize=\textwidth \epsfbox{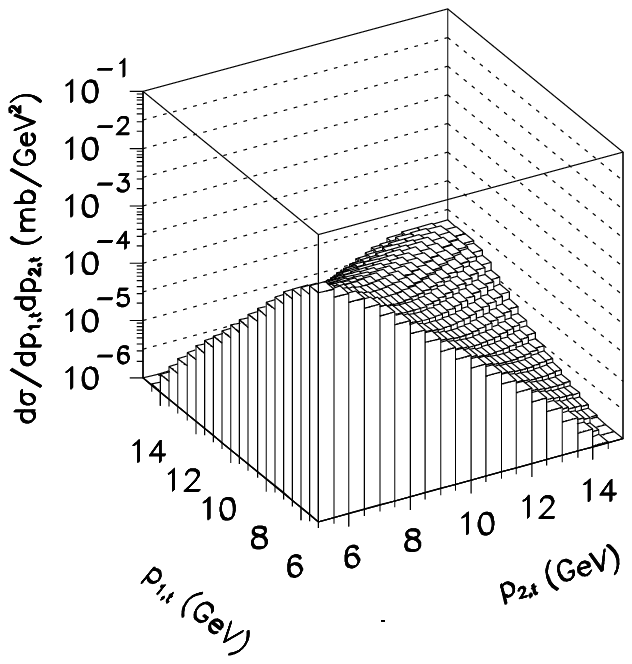}
\end{minipage}
\begin{minipage}{0.49\textwidth}
\epsfxsize=\textwidth \epsfbox{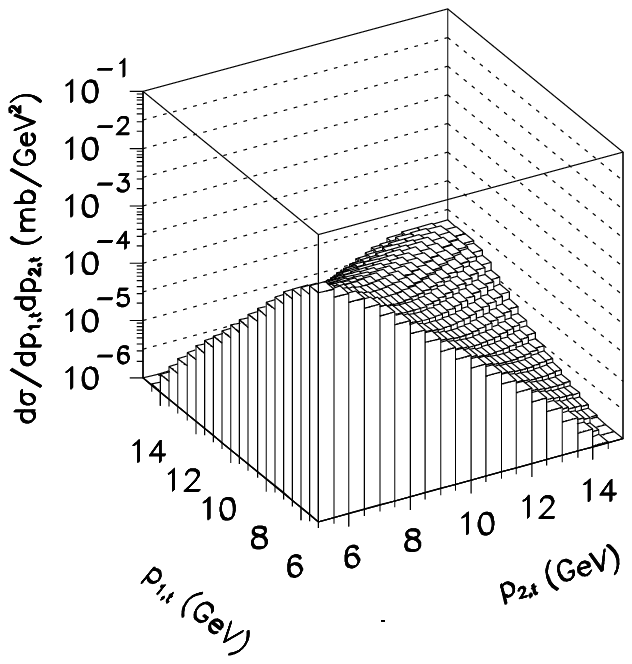}
\end{minipage}
\caption{\small
Two-dimensional distributions in $p_{1,t}$ and $p_{2,t}$ 
for different subprocesses $gg \to gg$ (left upper)
$gg \to q \bar q$ (right upper), $gq \to gq$ (left lower)
and $qg \to qg$ (right lower). In this calculation W = 200 GeV and
Kwieci\'nski UPDFs with exponential nonperturbative form factor
($b_0$ = 1 GeV$^{-1}$) and $\mu^2$ = 100 GeV$^2$ were used.
Here integration over full range of parton rapidities was made.
\label{fig:p1tp2t_kwiec_components}
}
\end{figure}


In Fig.\ref{fig:p1tp2t_kwiec_components} we show two-dimensional
maps in $(p_{1,t},p_{2,t})$ for listed above subprocesses.
Only very few approaches in the literature include both gluons and
quarks and antiquarks.
In the calculation above we have used Kwieci\'nski UPDFs with
exponential nonperturbative form factor ($b_0$ = 1 GeV$^{-1}$)
and the factorization scale $\mu^2 = (p_{t,min}+p_{t,max})^2/4$ = 100 GeV$^2$.

\begin{figure}[!hp]      
\begin{minipage}{0.49\textwidth}
\epsfxsize=\textwidth\epsfbox{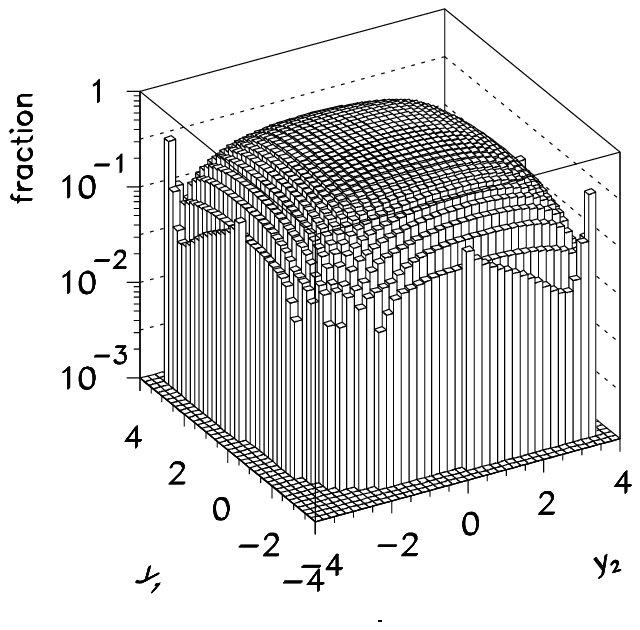}
\end{minipage}
\begin{minipage}{0.49\textwidth}
\epsfxsize=\textwidth \epsfbox{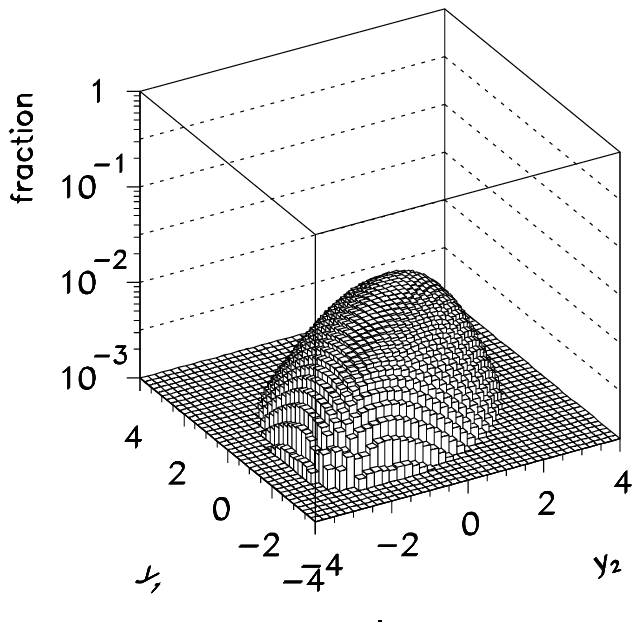}
\end{minipage}
\begin{minipage}{0.49\textwidth}
\epsfxsize=\textwidth \epsfbox{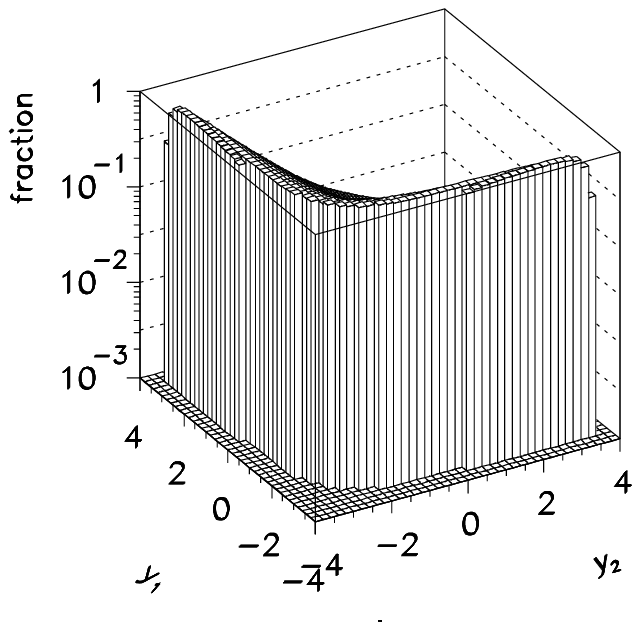}
\end{minipage}
\begin{minipage}{0.49\textwidth}
\epsfxsize=\textwidth \epsfbox{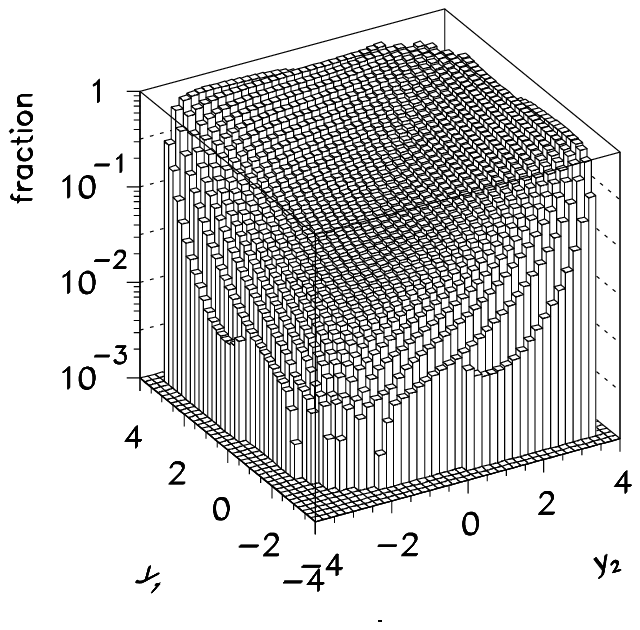}
\end{minipage}
\caption{\small
Two-dimensional distributions of fractional contributions of different
subprocesses
as a function of $y_1$ and $y_{2}$ 
for $gg \to gg$ (left upper)
$gg \to q \bar q$ (right upper), $gq \to gq$ (left lower)
and $qg \to qg$ (right lower). In this calculation W = 200 GeV and
Kwieci\'nski UPDFs with exponential nonperturbative form factor
and $b_0$ = 1 GeV$^{-1}$ were used. The integration is made
for jets from the transverse momentum interval:
5 GeV $< p_{1,t}, p_{2,t} <$ 20 GeV.
\label{fig:2to2_contributions_y1y2}}
\end{figure}


In Fig.\ref{fig:2to2_contributions_y1y2} we show a fractional
contributions (individual component to the sum of all four components)
of the above four processes on the two-dimensional map $(y_1,y_2)$.
One point here requires a better clarification.
Experimentally it is not possible to distinguish gluon and
quark/antiquark jets. Therefore in our calculation of the $(y_1,y_2)$
dependence one has to symmetrize the cross section (not the amplitude)
with respect to gluon -- quark/antiquark exchange 
($y_1 \to y_2, y_2 \to y_1$).
This can be done technically by exchanging $\hat t$ and $\hat u$
variables in the matrix element squared.
While at midrapidities the contribution of diagram $B_1$ + $B_2$
is comparable to the diagram $A_1$, at larger rapidities the
contributions of diagrams of the type B dominate. 
The contribution of diagram $A_2$ is relatively
small in the whole phase space. When calculating the contributions of
the diagram $A_1$ and $A_2$ one has to be careful about collinear
singularity which leads to a significant enhancement of the cross section
at $\phi_{-}$=0 and $y_1 = y_2$, i.e. in the one jet case.
This is particularly important for the matrix elements obtained by
the naive analytic continuation from the formula for on-shell initial partons.
The effect can be, however, easily eliminated with the jet-cone separation
algorithm discussed in Appendix D.

\begin{figure}[!h]   
 \centerline{\includegraphics[width=0.60\textwidth]{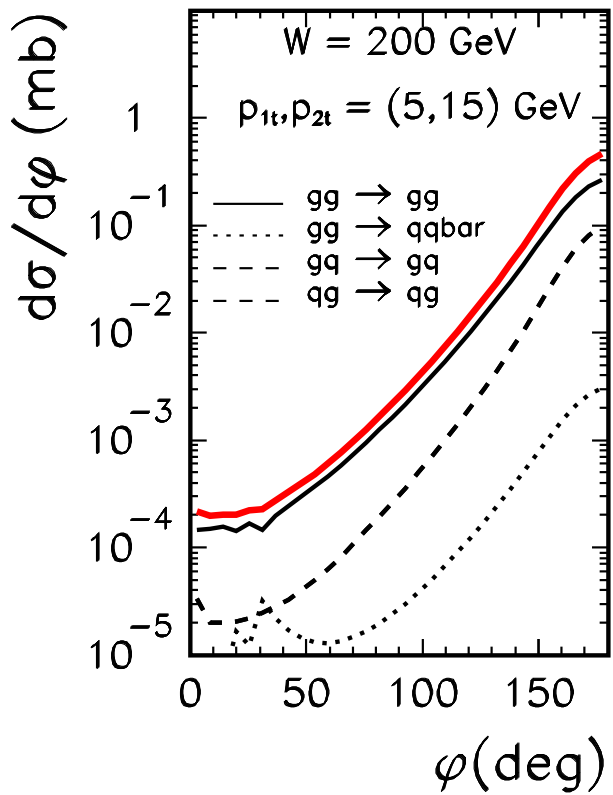}}
   \caption{ \label{fig:2to2_contributions_phi} 
\small 
The angular correlations for all four components: $gg \to gg$ (solid),
$gg \to q \bar q$ (dashed) and $gq \to gq$ = $qg \to qg$ (dash-dotted).
The calculation is performed with the Kwieci\'nski UPDFs and
$b_0$ = 1 GeV$^{-1}$.
The integration is made for jets from the transverse momentum interval:
5 GeV $< p_{1,t}, p_{2,t} <$ 15 GeV and from the rapidity interval:
-4 $< y_1, y_2 <$ 4.
}
\end{figure}


For completeness in Fig.\ref{fig:2to2_contributions_phi} we show
azimuthal angle dependence of the cross section for all four components.
There is no sizeable difference in the shape of azimuthal distribution
for different components.
 
\begin{figure}[!h]   
 \centerline{\includegraphics[width=0.6\textwidth]{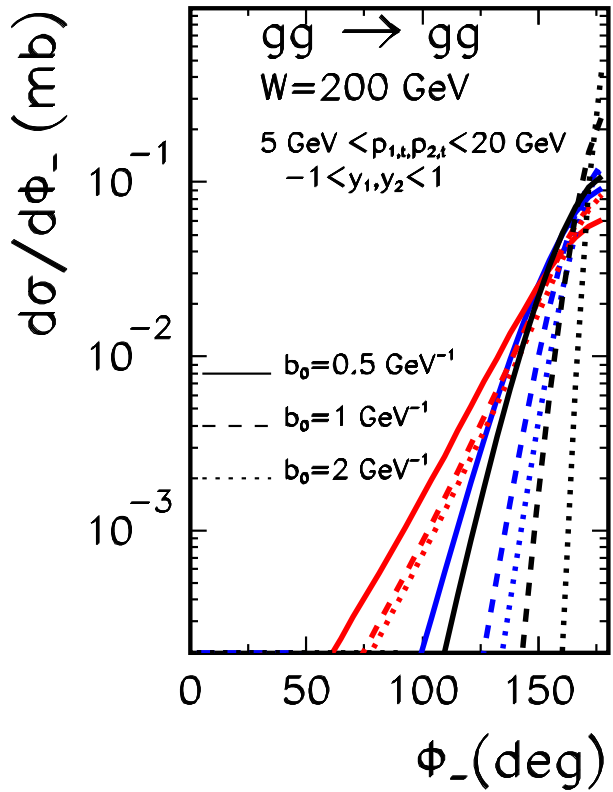}}
   \caption{ \label{fig:b0_mu2_phi}
\small
The azimuthal correlations for the $gg \to gg$ component obtained with
the Kwieci\'nski UGDFs for different values of the nonperturbative
parameter $b_0$ and for different evolution scales $\mu^2$ = 10 (on line
blue), 100 (on line red) GeV$^2$.
The initial distributions (without evolution) are shown for reference
by black lines.
}
\end{figure}

The Kwieci\'nski approach allows to separate the unknown perturbative
effects incorporated via nonperturbative form factors
and the genuine effects of QCD evolution.
The Kwieci\'nski distributions have two external parameters:
\begin{itemize}
\item the parameter $b_0$ responsible for nonperturbative effects,
such as primordial distribution of partons in the nucleon,
\item the evolution scale $\mu_F^2$ responsible for the soft
resummation effects.
\end{itemize}
While the latter can be identified physically with characteristic
kinematical quantities in the process $\mu_F^2 \sim p_{1,t}^2, p_{2,t}^2$,
the first one is of nonperturbative origin and cannot be calculated
from first principles.
The shapes of distributions depends, however, strongly on the value of
the parameter $b_0$.
This is demonstrated in Fig.\ref{fig:b0_mu2_phi} for the $gg \to gg$
subprocess.
The smaller $b_0$ the bigger decorrelation in azimuthal angle
can be observed. In Fig.\ref{fig:b0_mu2_phi} we show also the role of
the evolution scale in the Kwieci\'nski distributions.
The QCD evolution embedded in the Kwieci\'nski evolution equations
populate larger transverse momenta of partons entering the hard process.
This significantly increases the initial (nonperturbative) decorrelation
in azimuth.
For transverse momenta of the order of $\sim$ 10 GeV the effect of
evolution is of the same order of magnitude as the effect due to
nonperturbative physics. For larger scales of the order of $\mu_F^2
\sim$ 100 GeV$^2$, more adequate for jet production, the initial
condition is of minor importance and the effect of decorrelation
is dominated by the evolution. Asymptotically (infinite scales) there is
no dependence on the initial condition provided reasonable initial
conditions are taken.

\begin{figure}[!h]   
 \centerline{\includegraphics[width=0.6\textwidth]{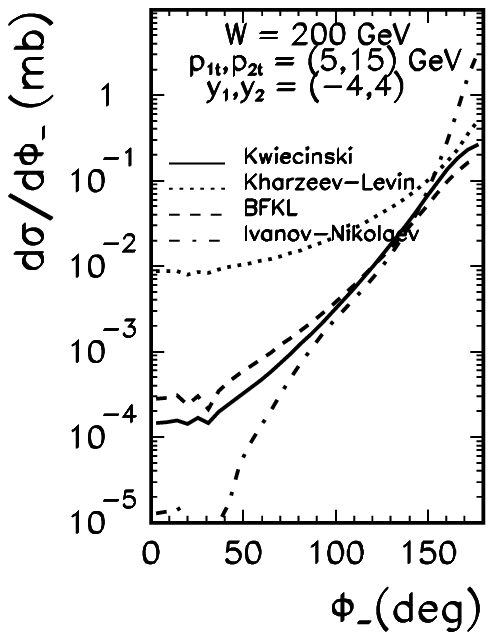}}
   \caption{ \label{fig:dsig_dphid_updf}
\small
The azimuthal correlations for the $gg \to gg$ component obtained 
for different UGDFs from the literature.
The Kwieci\'nski distribution is for $b_0$ = 1 GeV$^{-1}$ and $\mu^2$ =
100 GeV$^2$.
}
\end{figure}

In Fig.\ref{fig:dsig_dphid_updf} we show azimuthal-angle correlations
for the dominant at midrapidity $g g \to g g$ component
for different UGDFs from the literature. Rather different results are
obtained for different UGDFs. In principle, experimental results
could select the ``best'' UGDF. We do not need to mention that such
measurements are not easy at RHIC and rather hadron correlations are studied
instead of jet correlations.

\begin{figure}[!h]   
 \centerline{\includegraphics[width=0.45\textwidth]{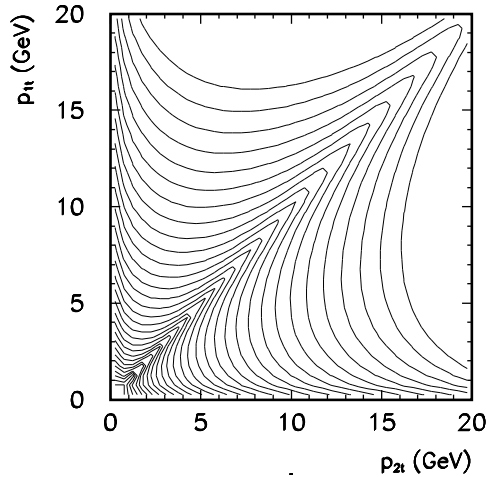}}
   \caption{ \label{fig:soft_regions}
\small 
Two-jet correlations for a $2 \to 3$ $gg \to ggg$ component
for RHIC energy W = 200 GeV. The soft singularities are shown as ridges.
The pQCD calculations are reliable outside of the regions of ridges.
}
\end{figure}


Before we start presenting further more detailed results let us
concentrate on NLO calculation
\footnote{Please note that what we call here NLO, is called sometimes
LO in the context of jet correlations \cite{D0_data}.}.
In Fig.\ref{fig:soft_regions} we show the results of a naive calculation,
on the $(p_{1,t},p_{2,t})$ plane where soft divergences are shown explicitly.
One clearly sees 3 sharp ridges: along x and y axes as well as along the
diagonal. While the ridges along x and y axis can be easily eliminated
by imposing cuts on $p_{1,t}$ and $p_{2,t}$, i.e. on jets taken
in the analysis of correlations. The elimination of the third ridge
is more subtle and will be discussed somewhat later.
Sometimes asymmetric cuts on jet transverse momenta are imposed in
order to avoid technical problems. 

\begin{figure}[!hp]      
\begin{minipage}{0.49\textwidth}
\epsfxsize=\textwidth\epsfbox{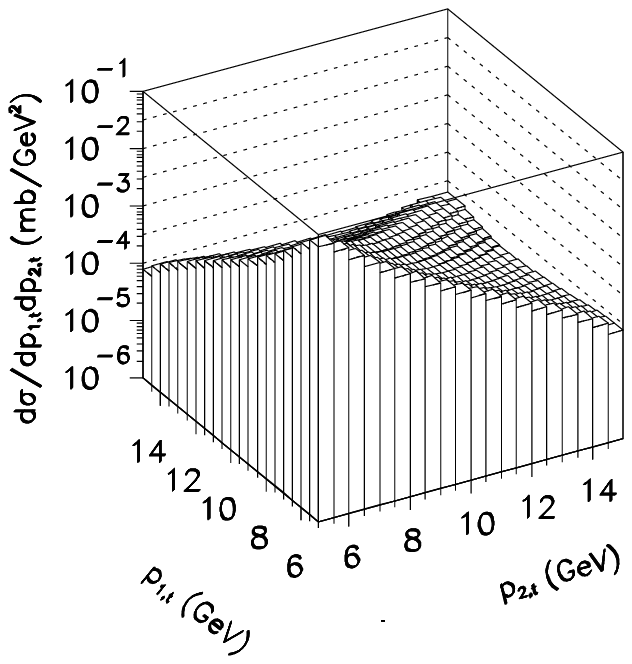}
\end{minipage}
\begin{minipage}{0.49\textwidth}
\epsfxsize=\textwidth \epsfbox{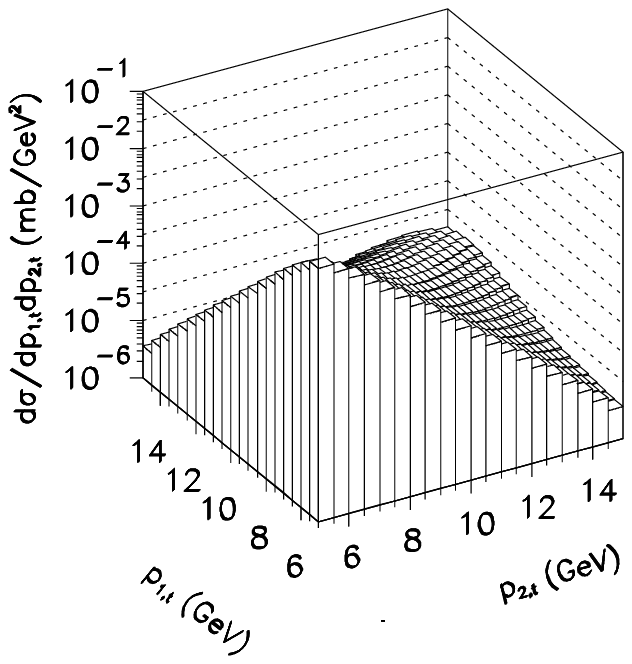}
\end{minipage}
\begin{minipage}{0.49\textwidth}
\epsfxsize=\textwidth \epsfbox{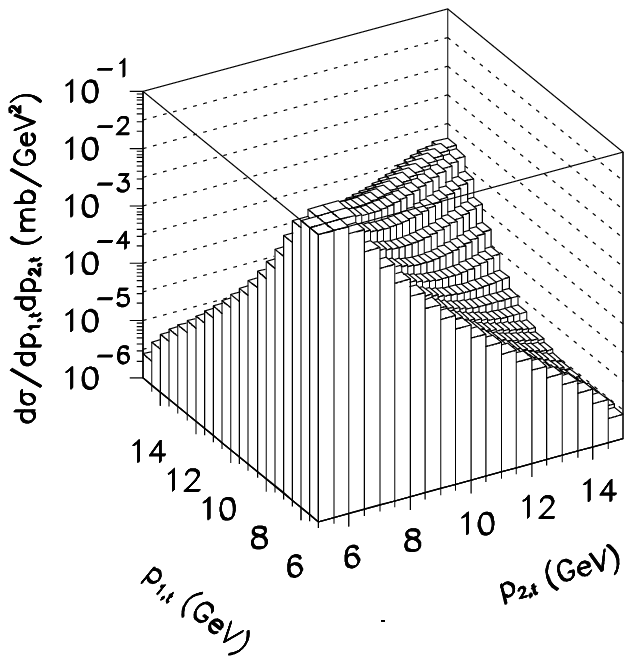}
\end{minipage}
\begin{minipage}{0.49\textwidth}
\epsfxsize=\textwidth \epsfbox{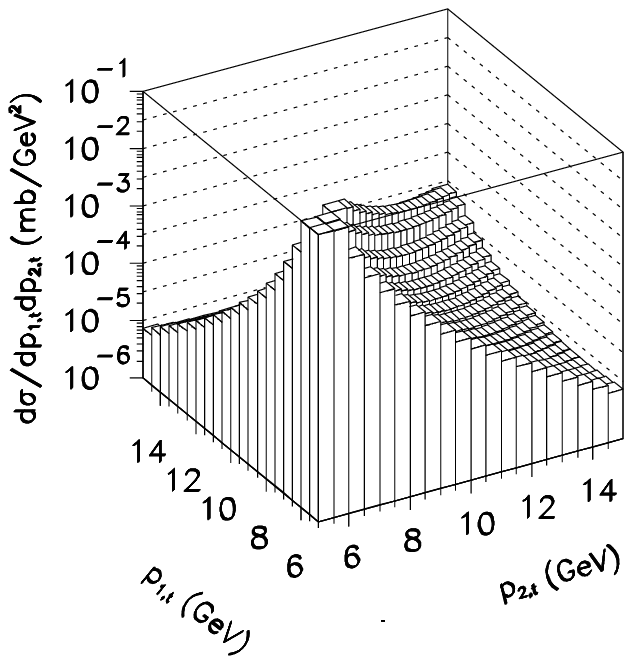}
\end{minipage}
*
\caption{\small
Two-dimensional distributions in $p_{1t}$ and $p_{2t}$ for
KL (left upper), BFKL (right upper), Ivanov-Nikolaev (left lower) UGDFs
and for the $g g \to g g g$ (right lower). In this calculation
-4 $< y_1, y_2<$ 4.
\label{fig:p1tp2t_maps}}
\end{figure}

Let us start from presenting the results on the plane $(p_{1,t},p_{2,t})$.
In Fig.\ref{fig:p1tp2t_maps} we show the maps for different choices
of UGDFs and for $2 \to 3$ processes in the broad range of transverse
momenta 5 GeV $< p_{1,t},p_{2,t} <$ 20 GeV for the RHIC energy W = 200
GeV. In this calculation we have not imposed any particular cuts
on rapidities. We have not imposed also any cut on the transverse momentum
of the unobserved third jet in the case of $2 \to 3$ calculation.
The small transverse momenta of the third jet contribute to the sharp
ridge along the diagonal $p_{1,t} = p_{2,t}$.
Naturally this is therefore very difficult to distinguish these three-parton
states from standard two jet events.
In principle, the ridge can be eliminated by imposing a cut on the
transverse momentum of the third (unobserved) parton.
There are also other methods to eliminate the ridge and underlying soft
processes which will be discussed somewhat later.

In Fig.\ref{fig:dsig_dphid_updf} we show corresponding distributions in
azimuthal angle $\phi_{-}$. Very different azimuthal correlation
functions are obtained for different UGDFs. The NLO azimuthal angle
correlation function exceeds those obtained in the $k_t$-factorization
approach for $\phi_{-} <$ 90$^o$. 

When calculating dijet correlations
in the standard NLO ($2 \to 3$) approach we have taken all possible dijet
combinations. This is different from what is usually taken in
experiments \cite{D0_data}, where correlation between leading jets
are studied. In our notation this means $p_{3,t} < p_{1,t}$ and
$p_{3,t} < p_{2,t}$. When imposing such extra condition on our NLO
calculation we get the dash-dotted curve in
Fig.\ref{fig:dsig_dphid_updf}.
In this case $d \sigma / d \phi_{-} = 0$ for $\phi_{-} < \tfrac{2}{3} \pi$.
This vanishing of the cross section is of purely kinematical origin.
Since in the $k_t$-factorization calculation only two jets are explicit,
there is no such an effect in this case.
This means that the region of $\phi_{-} < \frac{2}{3} \pi$ should
be useful to test models of UGDFs.
For completeness in Fig.\ref{fig:map_p1tp2t_lj} we show a
two-dimensional plot $(p_{1,t},p_{2,t})$ with imposing the leading-jet
condition.
Surprisingly the leading-jet condition removes a big part of
the two-dimensional space. In particular, regions with $p_{2,t} > 2 p_{1,t}$ 
(NLO-forbidden region1) and
$p_{1,t} > 2 p_{2,t}$ (NLO-forbidden region2) cannot be populated via
$2 \to 3$ subprocess \footnote{In LO collinear approach the whole plane,
except of the diagonal $p_{1,t} = p_{2,t}$, is forbidden.}.
There are no such limitations for $2 \to 4$, $2 \to 5$ and even
higher-order processes.
Therefore measurements in ``NLO-forbidden'' regions of the
$(p_{1,t},p_{2,t})$ plane
would test higher-order terms of the standard collinear pQCD.
These are also regions where UGDFs can be tested, provided that not too
big transverse momenta of jets taken into the correlation in order
to assure the dominance of gluon-initiated processes (for larger
transverse momenta and/or forward/backward rapidities one has to include
also quark/antiquark initiated processes via unintegrated
quark/antiquark distributions).

Can we gain a new information correlating the space of azimuthal angle
$(\phi_{-})$ and the space spanned by the lengths of transverse momenta
$(p_{1,t},p_{2,t})$ ?
In particular, it is interesting how the jet azimuthal
correlations depend on a region of $(p_{1,t},p_{2,t})$. For this purpose
in Fig.\ref{fig:windows} we define several regions in $(p_{1,t},p_{2,t})$,
called windows, for easy reference in the following.
They have been named $A_{ij}$ for future easy notation.
In Fig.\ref{fig:dsig_dphid_regions} we show angular azimuthal
correlations for each of these regions separately. While at small
transverse momenta the cross section obtained with $2 \to 2$
$k_t$-factorization and $2 \to 3$ collinear-factorization approaches
are of similar order, at larger transverse momenta and far from the diagonal
$p_{1,t} = p_{2,t}$ the cross section is dominated by the genuine
next-to-leading order processes. In these regions
the standard higher-order collinear-factorization approach seems
to be the best, and probably the only, method to study dijet
azimuthal-angle correlations.

Cuts on $p_{1,t}$ and $p_{2,t}$ remove a big part of soft singularities,
leaving only region of $p_{1,t} \approx p_{2,t}$.
In order to eliminate the regions where the pQCD calculation does not
apply we suggest to exclude the region shown in diagram
\ref{fig:excluded_diagonal_region} which is equivalent to including
the following cuts on the lengths of transverse momenta of the jets
taken into account in the correlations:
\begin{equation}
| p_{1,t} - p_{2,t} | > \Delta_{s}
\label{scalar_cut}
\end{equation}
In Fig.\ref{fig:dsig_dphid_scalar_cuts} we show the distribution of the
cross section in azimuthal angle for different (scalar) cuts 
$\Delta_{s}$ = 0,2,5 GeV.
We have also tried another way to remove singularities:
\begin{equation}
| \vec{p}_{1,t} + \vec{p}_{2,t} | > \Delta_{v}
\label{vector_cut}
\end{equation}
In Fig.\ref{fig:dsig_dphid_vector_cuts} we show the distribution of the
cross section in azimuthal angle for different (vector) cuts 
$\Delta_{v}$ = 0,2,5 GeV. These results are very similar to those obtained
with scalar cuts.

Both scalar and vector cuts remove efficiently the singularity
of the collinear 2$\to$3 contribution at $\phi_{-} = \pi$.
If too big values of $\Delta_s$ or $\Delta_v$ are used the cross section
of the $k_t$-factorization 2$\to$2 contribution is reduced considerably.

\section{Discussion and Conclusions}

Motivated by the recent experimental results of hadron-hadron correlations
at RHIC we have discussed dijet correlations in proton-proton collisions.
We have considered and compared results obtained
with collinear next-to-leading order approach and leading-order
$k_t$-factorization approach. 

In comparison to recent works in the framework of $k_t$-factorization
approach, we have included two new mechanisms based on $gq \to gq$
and $qg \to qg$ hard subprocesses.
This was done based on the Kwieci\'nski unintegrated parton
distributions.
We find that the new terms give significant contribution at RHIC energies.
In general, the results of the $k_t$-factorization approach depend
on UGDFs/UPDFs used, i.e. on approximation and assumptions made
in their derivation.

An interesting observation has been made for azimuthal angle correlations.
At relatively small transverse momenta ($p_t \sim$ 5--10 GeV)
the $2 \to 2$ subprocesses, not contributing to the correlation
function in the collinear approach, dominate over $2 \to 3$ components.
The latter dominate only at larger transverse momenta, i.e. in
the traditional jet region.

The results obtained in the standard NLO approach depend significantly
whether we consider correlations of any jets or correlations of only
leading jets. 
In the NLO approach one obtains
$\frac{d \sigma}{d \phi_{-}}$ = 0 if $\phi_{-} < \tfrac{2}{3} \pi$
for leading jets as a result of a kinematical constraint.
Similarly $\frac{d \sigma}{d p_{1,t} d p_{2,t}}$ = 0 if $p_{1,t} > 2
p_{2,t}$ or $p_{2,t} > 2 p_{1,t}$.

There is no such a constraint in the $k_t$-factorization approach
which gives a nonvanishing cross section at small relative azimuthal
angles between leading jets and transverse-momentum asymmetric
configurations. We conclude that in these regions the $k_t$-factorization
approach is a good and efficient tool for the description of leading-jet
correlations.
Rather different results are obtained with different UGDFs
which opens a possibility to verify them experimentally.
Alternatively, the NLO-forbidden configurations can be described
only by higher-order (NNLO and higher-order) terms.
We do not need to mention that this is a rather difficult and technically
involved computation.

On the contrary, in the case of correlations of any unrestricted jets
(all possible dijet combinations)
the NLO cross section exceeds the cross section obtained in
the $k_t$-factorization approach with different UGDFs. This is therefore
a domain of the standard fixed-order pQCD.
We recommend such an analysis as an alternative to study leading-jet
correlations. In principle, such an analysis could be done for
the already collected Tevatron data.

What are consequences for particle-particle correlations measured
recently at RHIC requires a separate dedicated analysis.
Here the so-called leading particles may come both from leading
and non-leading jets.
This requires taking into account the jet fragmentation process.
We leave this analysis for a separate study.

\vskip 0.5cm

\section{Appendices}

\subsection{Matrix elements for $2 \to 2$ processes
with initial off-shell gluons}

In this paper we shall include the following $2 \to 2$ processes
with at least one gluon in the initial state:\\
(a) $gg \to gg$, (b) $gg \to q \bar q$, (c) $gq \to gq$, (d) $qg \to qg$,
i.e. processes giving significant contributions for inclusive
jet production at relatively small jet transverse momenta and
midrapidities \cite{SB98}.
The last two processes were not included in Refs.\cite{LO00}, \cite{Bartels}.
We shall show that at RHIC energies they give contributions similar
(or even larger) to the contribution of the asymptotically dominant
$gg \to gg$ subprocess.

The matrix elements for on-shell initial gluons/partons read 
(see e.g.\cite{BP_book})
\begin{eqnarray}
\overline{|{\cal M}_{gg \to gg}|^2} &=&
\frac{9}{2} g_s^4 \left(
3 - \frac{\hat{t} \hat{u}}{\hat{s}^2} 
  - \frac{\hat{s} \hat{u}}{\hat{t}^2}
  - \frac{\hat{s} \hat{t}}{\hat{u}^2}
\right) \; ,
\nonumber \\
\overline{|{\cal M}_{gg \to q\bar q}|^2} &=&
\frac{1}{8} g_s^4 \left(
6 \frac{\hat{t}\hat{u}}{\hat{s}^2}
+\frac{4}{3} \frac{\hat{u}}{\hat{t}}
+\frac{4}{3} \frac{\hat{t}}{\hat{u}}
+3 \frac{\hat{t}}{\hat{s}}
+3 \frac{\hat{u}}{\hat{s}}   
\right) \; ,
\nonumber \\
\overline{|{\cal M}_{gq \to gq}|^2} &=&
g_s^4 \left(
-\frac{4}{9} \frac{\hat{s}^2+\hat{u}^2}{\hat{s}\hat{u}}
+ \frac{\hat{u}^2+\hat{s}^2}{\hat{t}^2}
\right) \;,
\nonumber \\
\overline{|{\cal M}_{qg \to qg}|^2} &=&
g_s^4 \left(
-\frac{4}{9} \frac{\hat{s}^2+\hat{t}^2}{\hat{s}\hat{t}}
+ \frac{\hat{t}^2+\hat{s}^2}{\hat{u}^2}
\right) \;. 
\end{eqnarray}
For on-shell initial gluons (partons) $\hat{s}+\hat{t}+\hat{u}=0$.

The matrix elements for off-shell initial gluons are obtained
by using the same formulas but with $\hat{s}, \hat{t}, \hat{u}$
calculated including off-shell initial kinematics.
In this case $\hat{s}+\hat{t}+\hat{u}=k_1^2+k_2^2$, where
$k_1^2, k_2^2 <$ 0 are virtualities of the initial gluons.
Our prescription can be treated as a smooth analytic continuation
of the on-shell formula off mass shell.
With our choice of initial gluon four-momenta $k_1^2 = -k_{1,t}^2$
and $k_2^2 = -k_{2,t}^2$.

In Refs.\cite{LO00,O00} another formula which includes off-shellness
of initial gluons was presented
\begin{equation}
\frac{d \sigma}{d^2p_{1,t} d^2p_{2,t} dy_1 dy_2} =
\int \frac{d^2 k_{1,t}}{\pi} \frac{d^2 k_{2,t}}{\pi}
{\cal F}(x_1,k_{1t}^2) 
\frac{d \sigma}{d^2p_{1,t} d^2p_{2,t}}
{\cal F}(x_2,k_{2,t}^2)   \; ,
\label{lo_cross_section}
\end{equation}
where 
\begin{equation}
\frac{d \sigma}{d^2p_{1,t} d^2p_{2,t}} =
2 \frac{N_c^2}{(N_c^2-1)} \; \alpha_s^2(\mu_r) \; \frac{1}{k_{1,t}^2 k_{2,t}^2}
\; \delta^2(\vec{k}_{1,t} + \vec{k}_{2,t} - \vec{p}_{1,t} - \vec{p}_{2,t} )
\; {\cal A} \; .
\label{LO_elementary_cross_section}
\end{equation}
The factor ${\cal A}$ is a function of momenta entering the hard process
${\cal A} = {\cal A}(\hat{s},\hat{t},\hat{u},k_{1,t},k_{2,t})$ (see
\cite{LO00}). The factor {\cal A} has been rederived recently in Ref.
\cite{Bartels} and the result of Leonidov and Ostrovsky was confirmed.

Please note a different convention of UGDF in our paper (${\cal F}$) with
those in Refs.\cite{LO00,O00} ($f$). The UGDFs in the two conventions
are related to each other as
\begin{equation}
{\cal F}(x,k_t^2) = f(x,k_t^2)/k_t^2  \; .
\end{equation}
In order to eliminate the delta function in 
Eq.(\ref{LO_elementary_cross_section}) we can
use the same tricks as in the previous section.

The formula of Leonidov and Ostrovsky is equivalent to our formula
if we define:
\begin{equation}
\overline{|{\cal M}|^2}_{off-shell} = 16 \pi^2 (x_1 x_2 s)^2
\frac{N_c^2-1}{2 N_c^2} \alpha_s^2 \frac{\cal A}{k_{1,t}^2 k_{2,t}^2} \; .
\end{equation}
%


\subsection{Matrix elements for $2 \to 3$ processes}

In this subsection we list the squared matrix elements
averaged and summed over initial and final spins and colors
used to calculate the contribution of the $2 \to 3$ partonic processes
(For useful reference see e.g.\cite{berends,BP_book}).

For the $gg \to ggg$ process ($k_1 + k_2 \to k_3+k_4+k_5$) 
the squared matrix element is
\begin{equation}
\begin{split}
\overline{ | {\cal M} |^2 } & =
\frac{1}{2} g_s^6 \frac{N_c^3}{N_c^2-1}  \\
\bigl[
&(12345)+(12354)+(12435)+(12453)+(12534)+(12543)+  \\ 
&(13245)+(13254)+(13425)+(13524)+(12453)+(14325) 
\bigr] \\
&\times \sum_{i<j} (k_i k_j) / \prod_{i<j} (k_i k_j) \; ,
\label{ME_gg_ggg}
\end{split}
\end{equation}
where $(ijlmn) \equiv (k_i k_j) (k_j k_l) (k_l k_m) (k_m k_n) (k_n k_i)$. 

It is useful to calculate matrix element for the process
$q \bar q \to g g g$.
The squared matrix elements for other processes can be obtained
by crossing the squared matrix element for the process
$q \bar q \to g g g$ ($p_a + p_b \to k_1 + k_2 + k_3$)
\begin{equation}
\begin{split}
\overline{ | {\cal M} |^2 } & =
g_s^6 \frac{N_c^2-1}{4 N_c^4}  \\
& \sum_{i}^{3} a_i b_i (a_i^2+b_i^2) / (a_1 a_2 a_3 b_1 b_2 b_3)
\\ \times
&\Biggl[
\frac{\hat{s}}{2} + N_c^2 \left(
\frac{\hat{s}}{2}
-\frac{a_1 b_2 + a_2 b_1}{(k_1 k_2)}
-\frac{a_2 b_3 + a_3 b_2}{(k_2 k_3)}
-\frac{a_3 b_1 + a_1 b_3}{(k_3 k_1)}
\right)  \\
&+ \frac{2 N^4}{\hat{s}}
\left(
\frac{a_3 b_3 (a_1 b_2 + a_2 b_1)}{(k_2 k_3)(k_3 k_1)} +
\frac{a_1 b_1 (a_2 b_3 + a_3 b_2)}{(k_3 k_1)(k_1 k_2)} +
\frac{a_2 b_2 (a_3 b_1 + a_1 b_3)}{(k_1 k_2)(k_2 k_3)} 
\right)
\Biggr] \; ,
\end{split}
\end{equation}
where the quantities $a_i$ and $b_i$ are defined as:
\begin{eqnarray}
a_i \equiv (p_a k_i) \; , \nonumber \\
b_i \equiv (p_b k_i) \; .
\label{auxiliary_ai_bi}
\end{eqnarray}

The matrix element for the process $gg \to q\bar q g$ is obtained from
that of $q \bar q \to ggg$ by appropriate crossing:
\begin{equation}
\overline{| {\cal M} |^2}_{gg \to q\bar q g}(k_1,k_2,k_3,k_4,k_5)=
\frac{9}{64} \cdot
\overline{| {\cal M} |^2}_{q \bar q \to ggg}(-k_4,-k_3,-k_1,-k_2,k_5) 
\; .
\label{crossing1}
\end{equation}
We sum over 3 final flavours (f = u, d, s).

For the $qg \to qgg$ process
\begin{equation}
\overline{| {\cal M} |^2}_{qg \to q g g}(k_1,k_2,k_3,k_4,k_5)=
\left(-\frac{3}{8}\right) \cdot
\overline{| {\cal M} |^2}_{q \bar q \to ggg}(k_1,-k_3,-k_2,k_4,k_5)
\label{crossing2a}
\end{equation}
and finally for the process $g\bar q \to \bar q g g$
\begin{equation}
\overline{| {\cal M} |^2}_{g\bar q \to \bar q g g}(k_1,k_2,k_3,k_4,k_5)=
\left(-\frac{3}{8}\right) \cdot
\overline{| {\cal M} |^2}_{q \bar q \to ggg}(-k_3,k_2,-k_1,k_4,k_5) \; .
\label{crossing2b}
\end{equation}

The squared matrix elements are used then in formula
(\ref{2to3_parton_formula}). The contributions with two quark/antiquark
initiated processes are important at extremely large rapidities.
They will be neglected in the present analysis where we concentrate
on midrapidities.

\subsection{Running $\alpha_s$}

The treatment of the running coupling constants in $2 \to 2$ and $2 \to 3$
subprocesses is important in numerical evaluation of the cross
section.

For the $2 \to 2$ case we shall try several prescriptions:\\
($\alpha_1$) $\alpha_s^2 = \alpha_s(p_{1,t}^2) \alpha_s(p_{2,t}^2)$, \\
($\alpha_2$) $\alpha_s^2 = \alpha_s^2(\frac{p_{1,t}^2+p_{2,t}^2}{2})$, \\
($\alpha_3$) $\alpha_s^2 = \alpha_s^2(p_{1,t} p_{2,t})$.

Analogously for the $2 \to 3$ case:\\
($\beta_1$) $\alpha_s^2 = 
\alpha_s(p_{1t}^2) \alpha_s(p_{2,t}^2) \alpha_s(p_{3,t}^2)$, \\
($\beta_2$) $\alpha_s^2 = 
\alpha_s^3(\frac{p_{1,t}^2+p_{2,t}^2+p_{3,t}^2}{3})$.


\subsection{Jet separation}

In order to make reference to real situation, as in experiments,
one has to take care about separation of jets in the azimuthal angle
and rapidity space.

In the case of $k_t$-factorization calculation, when there are
only two explicit jets we impose the following jet-cone condition:
\begin{equation}
R_{12} = \sqrt{ (\Delta \phi_{12})^2 + (y_1 - y_2)^2 } < R_0 \; .
\label{R_12_cone}
\end{equation}
Of course in this case $\Delta \phi_{12} = \phi_{-}$. $R_0$ is an
external parameter. For reasonable values of $R_0 <$ 1 the condition may be
active only for small $\phi_{-}$. We discuss 
the role of the extra cut in the paper.

In the case of $2 \to 3$ subprocesses one has to check two extra
conditions:
\begin{eqnarray}
R_{13} = \sqrt{ (\Delta \phi_{13})^2 + (y_1 - y_3)^2 } < R_0 \; ,
\nonumber \\
R_{23} = \sqrt{ (\Delta \phi_{23})^2 + (y_2 - y_3)^2 } < R_0 \; .
\label{R_13_R_23_cones}
\end{eqnarray}
%
Here one can expect slightly more complicated situation.
Those two cuts reduce the correlation function everywhere in
$\phi_{-}=\Delta\phi_{12}$.

{\bf Acknowledgments} 
We acknowledge the participation of Marta Tichoruk in the preliminary
stage of the analysis. We are very indebted to Tomasz Pietrycki
for help in preparing some more complicated figures.
The discussion with Wolfgang Sch\"afer is greatly acknowledged.
We are indebted to Andreas van Hameren for teaching us how to use the
computer package HELAC for multiparton production.
We are also indebted to Alexander Kupco and Markus Wobisch for explaining
some details of the measurement and calculations, respectively,
concerning the dijet production at the Tevatron.
This work was partially supported by the grant
of the Polish Ministry of Scientific Research and Information Technology
number 1 P03B 028 28.


\newpage

\begin{figure}[!h]   
 \centerline{\includegraphics[width=0.45\textwidth]{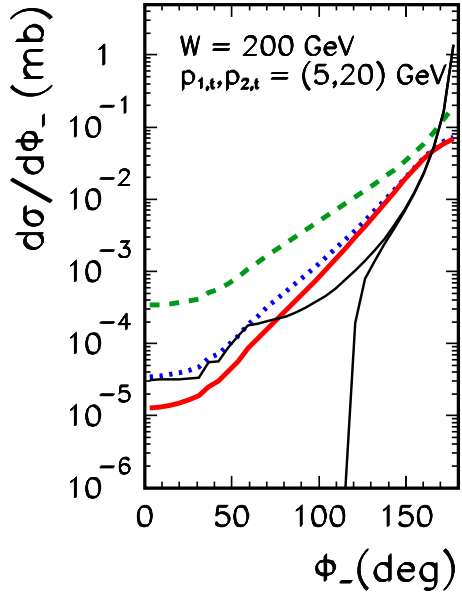}}
   \caption{ \label{fig:dsig_dphi}
\small 
Jet-jet azimuthal correlations $d \sigma / d\phi_{-}$
for the $gg \to gg$ component and different UGDFs
as a function of azimuthal angle between the gluonic jets. 
In this calculation W = 200 GeV and -1 $< y_1, y_2<$ 1,
5 GeV $< p_{1t}, p_{2t} <$ 20 GeV. The notation here is the same as in
Fig.\ref{fig:dsig_dphid_updf}.
}
\end{figure}

\begin{figure}[!h]
 \centerline{\includegraphics[width=0.45\textwidth]{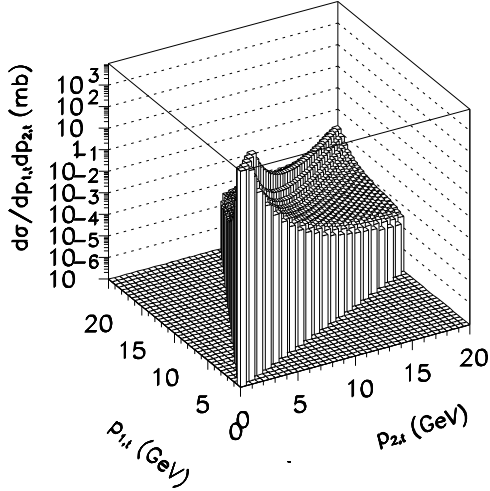}}
   \caption{ \label{fig:map_p1tp2t_lj}
\small
Cross section for the $gg \to ggg$ component on the $(p_{1,t},p_{2,t})$ plane
with the condition of leading jets (partons). The borders of NLO
accessible regions are clearly visible.
}
\end{figure}

\begin{figure}[!h]   
 \centerline{\includegraphics[width=0.45\textwidth]{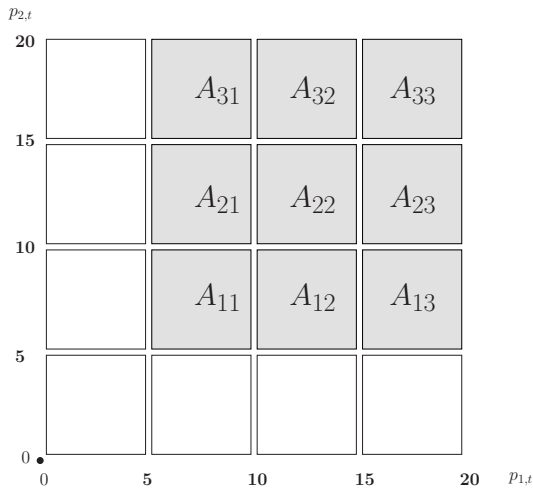}}
   \caption{ \label{fig:windows}
\small 
Definition of windows in $(p_{1,t},p_{2,t})$ plane for a further use.
}
\end{figure}

\begin{figure}[!h]   
 \centerline{\includegraphics[width=0.80\textwidth]{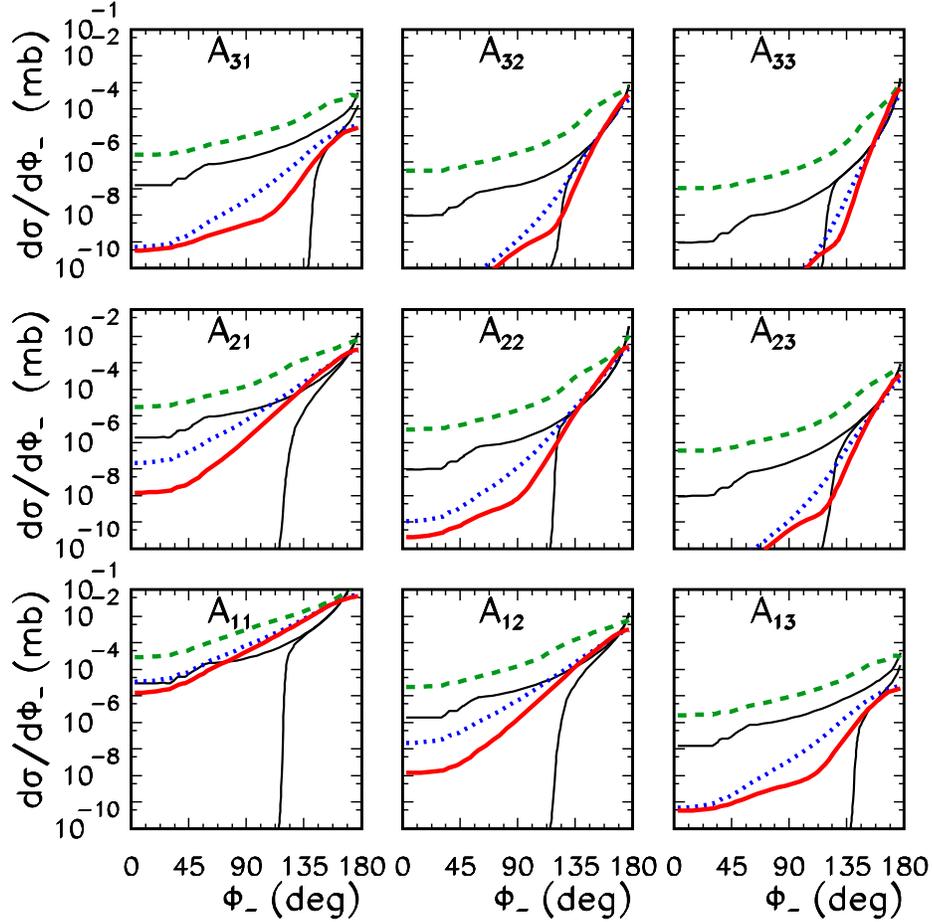}}
   \caption{ \label{fig:dsig_dphid_regions}
\small 
Dijet azimuthal correlations $d \sigma / d\phi_{-}$ for different windows in
the $(p_{1,t},p_{2,t})$ plane as a function of relative azimuthal angle
$\phi_{-}$ between outgoing jets for RHIC energy W = 200 GeV.
The jet-cone radius $R_{12}$ = 1 was used here in addition to separate
jets. The notation here is the same as in
Fig.\ref{fig:dsig_dphid_updf}.
}
\end{figure}

\begin{figure}[!h]   
 \centerline{\includegraphics[width=0.8\textwidth]{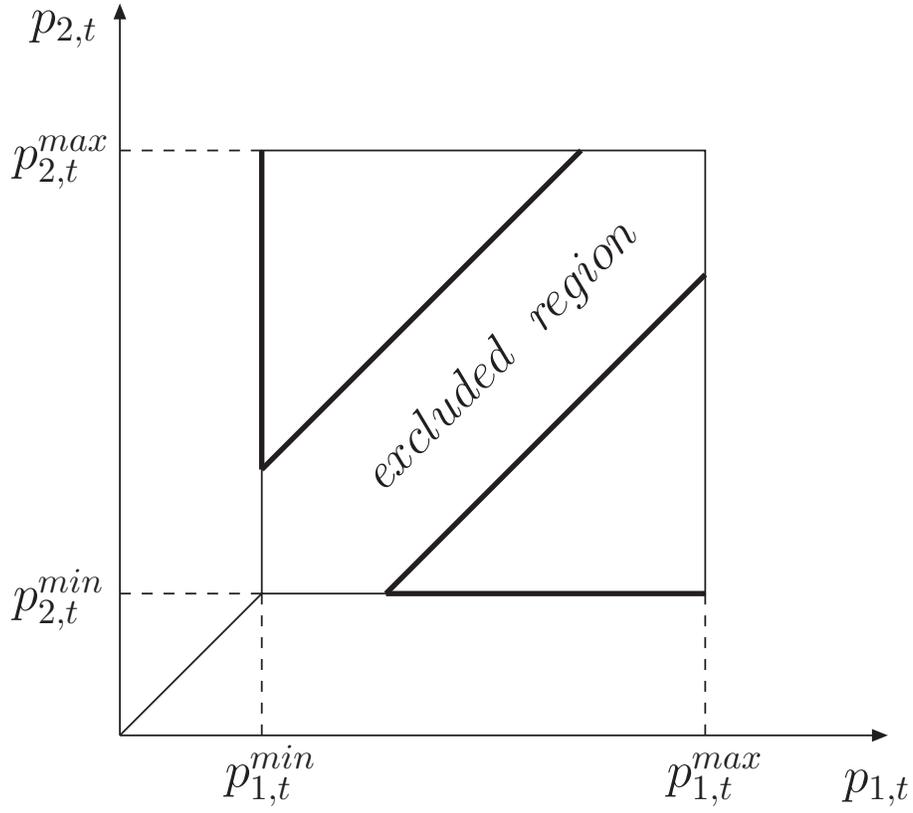}}
   \caption{ \label{fig:excluded_diagonal_region}
\small 
The excluded diagonal region.
Shown are also standard cuts on jet transverse momenta.
}
\end{figure}

\begin{figure}[!hp]      
\includegraphics[width=1.0\textwidth]{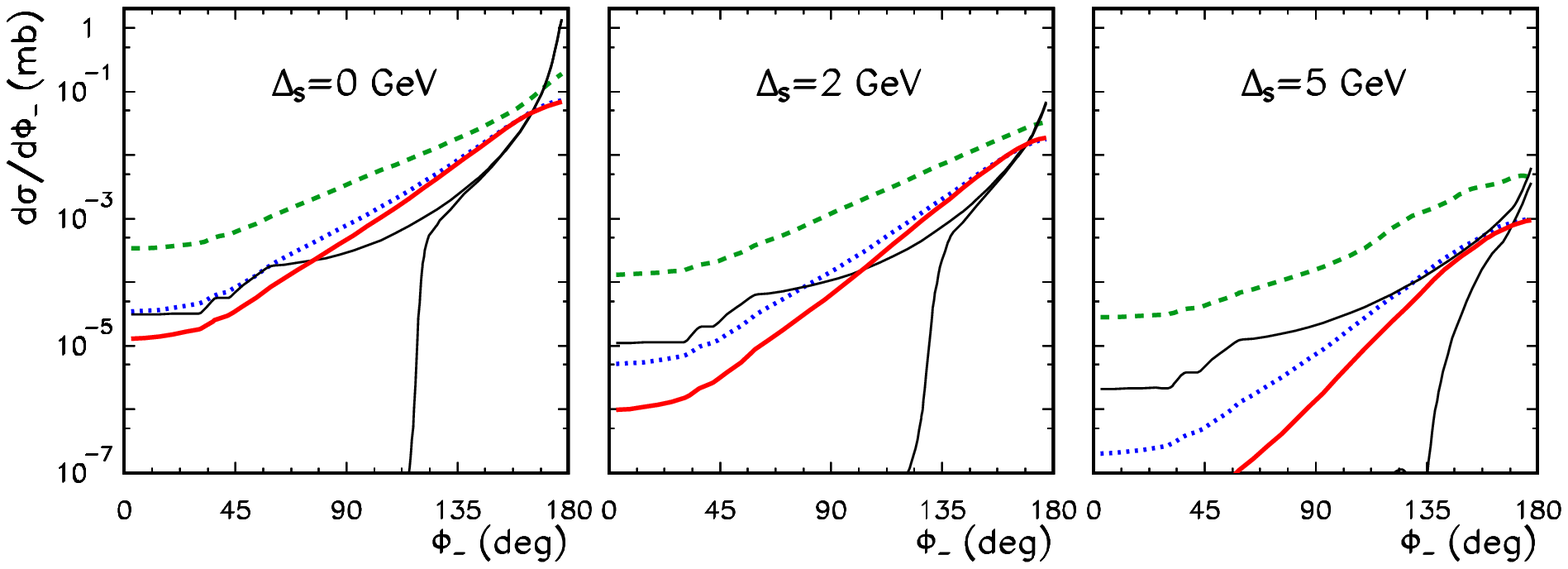}
\caption{\small
Azimuthal angular correlations for different values of the parameter
$\Delta_s$ = 0, 2, 5 GeV. Different UGDF are used. The notation
here is the same as previously. The jet-cone radius $R_{12}$ = 1 was
used in addition to separate jets. The notation here is the same as in
Fig.\ref{fig:dsig_dphid_updf}.
\label{fig:dsig_dphid_scalar_cuts}}
\end{figure}

\begin{figure}[!hp]      
\includegraphics[width=1.0\textwidth]{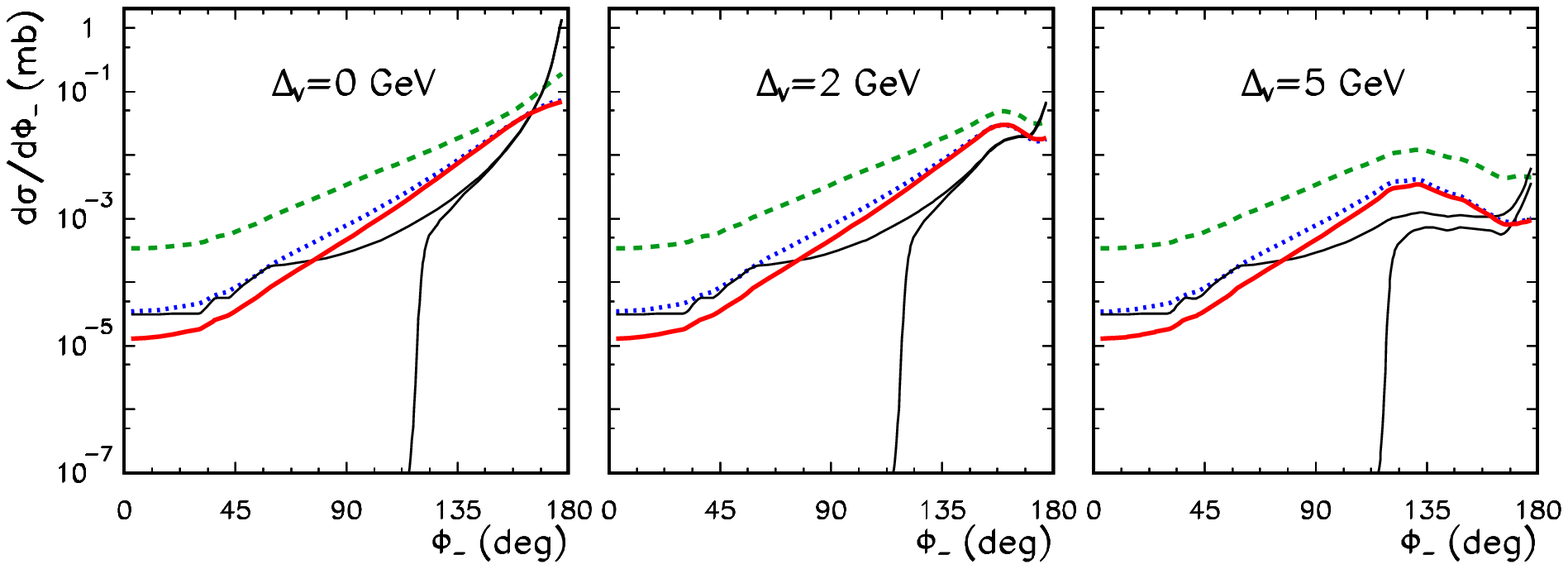}
\caption{\small
Azimuthal angular correlations for different values of the parameter
$\Delta_v$ = 0, 2, 5 GeV. Different UGDF are used. The notation
here is the same as previously. The jet-cone radius $R_{12}$ = 1 was
used in addition to separate jets. The notation here is the same as in
Fig.\ref{fig:dsig_dphid_updf}.
\label{fig:dsig_dphid_vector_cuts}}
\end{figure}




\begin{thebibliography}{100}

\bibitem{RHIC_correlations_nucleus_nucleus}
S.S. Adler et al. (PHENIX collaboration), Phys. Rev. Lett. {\bf 97}
(2006) 052301; \\
S.S. Adler et al. (PHENIX collaboration), Phys. Rev. {\bf C73} (2006)
054903;\\
S.S. Adler et al. (PHENIX collaboration), Phys. Rev. Lett. {\bf 96}
(2006) 222301;\\
M. Oldenburg et al. (STAR collaboration), Nucl. Phys. {\bf A774} (2006) 507.

\bibitem{RHIC_correlations_proton_proton}
S.S. Adler et al. (PHENIX collaborations), Phys. Rev. {\bf D74} (2006) 
072002. 

\bibitem{Levai}
P. Levai, G. Fai and G. Papp, Phys. Lett. {\bf B634} (2006) 383.

\bibitem{original_kt_factorization}
S. Catani, M. Ciafaloni and F. Hautmann, Nucl. Phys. {\bf 366} (1991)
135;\\
J.C. Collins and R.K. Ellis, Nucl. Phys. {\bf B360} (1991) 3.

\bibitem{UGDF_HERA}
J. Kwieci\'nski, A.D. Martin and A.M. Sta\'sto, 
Phys. Rev. {\bf D56} (1997) 3991; \\
I.P. Ivanov and N.N. Nikolaev,
Phys. Rev. {\bf D65} (2002) 054004;\\
H. Jung and G. Salam,
Eur. Phys. Jour. {\bf C19} (2002) 351.

\bibitem{BS00}
S.P. Baranov and M. Smi{\v z}anska, Phys. Rev. {\bf D62} (2000) 014012.

\bibitem{LSZ02}
A.V. Lipatov, V.A. Saleev and N.P. Zotov, hep-ph/0112114;\\
S.P. Baranov, A.V. Lipatov and N.P. Zotov, hep-ph/0302171,
Yad. Fiz. {\bf 67} (2004) 856.

\bibitem{LS04}
M. {\L}uszczak and A. Szczurek, Phys. Lett. {\bf B594} (2004) 291.

\bibitem{LS06}
M. {\L}uszczak and A. Szczurek, Phys. Rev. {\bf D73} (2006) 054028.

\bibitem{LZ05_photon}
A.V. Lipatov and N.P. Zotov, Phys. Rev. {\bf D72} (2005) 054002.

\bibitem{PS07}
T. Pietrycki and A. Szczurek, hep-ph/0606304,
Phys. Rev.{\bf D75} (2007) 014023.

\bibitem{Higgs}
A.V. Lipatov and N.P. Zotov, Eur. Phys. J. {\bf C44} (2005) 559;\\ 
M.~{\L}uszczak and A.~Szczurek, Eur. Phys. J. {\bf C46} 123 (2006).

\bibitem{KS04}
J. Kwieci\'nski and A. Szczurek, Nucl. Phys. {\bf B680} (2004) 164.

\bibitem{szczurek03}
A. Szczurek, Acta Phys. Polon. {\bf B34} (2003) 3191.

\bibitem{CS05}
M. Czech and A. Szczurek, Phys. Rev. {\bf C72} (2005) 015202;\\
M. Czech and A. Szczurek, J. Phys. {\bf G32} (2006) 1253.

\bibitem{SNSS01}
A. Szczurek, N.N. Nikolaev, W. Sch\"afer and J. Speth, Phys. Lett. {\bf
  B500} (2001) 254.

\bibitem{LO00}
A. Leonidov and D. Ostrovsky, Phys. Rev. {\bf D62} (2000) 094009.


\bibitem{SB98}
A. Szczurek and A. Budzanowski, Phys. Lett. {\bf B404} (1998) 141.

\bibitem{AM04}
U. D'Alesio and F. Murgia, Phys. Rev. {\bf D70} (2004) 074009.

\bibitem{BP_book}
V.D. Barger and R.J.N. Phillips, ``Collider physics'',
Addison-Wesley Publishing Company, 1987

\bibitem{berends}
F.A. Berends, R. Kleiss, P.De Causmaecker, R. Gastmans, and T.T. Wu,
Phys. Lett. {\bf B 103} (1981) 124.

\bibitem{O00}
D. Ostrovsky, Phys. Rev. {\bf D62} (2000) 054028.

\bibitem{kwiecinski}
J. Kwieci\'nski, Acta Phys. Polon. {\bf B33} (2002) 1809;\\
A. Gawron and J. Kwieci\'nski, Acta Phys. Polon. {\bf B34}
(2003) 133;\\
A. Gawron, J. Kwieci\'nski and W. Broniowski, Phys. Rev. {\bf D68}
(2003) 054001.


\bibitem{Bartels}
J. Bartels, A. Sabio Vera and F. Schwennsen, hep-ph/0608154, JHEP 0611
(2006) 051.


\bibitem{KMR}
M.A. Kimber, A.D. Martin and M.G. Ryskin,
Eur. Phys. J. {\bf C12} (2000) 655;\\
M.A. Kimber, A.D. Martin and M.G. Ryskin, Phys. Rev. {\bf D63}
(2001) 114027.

\bibitem{KMS97}
J. Kwieci\'nski, A.D. Martin and A.M. Sta\'sto, Phys. Rev. {\bf D56} (1997) 3991.

\bibitem{BFKL}
E.A. Kuraev, L.N. Lipatov and V.S. Fadin,
Sov. Phys. JETP {\bf 45} (1977) 199;\\
Ya.Ya. Balitskij and L.N. Lipatov, Sov. J. Nucl. Phys. {\bf 28} (1978)
822.

\bibitem{AKMS94}
A.J. Askew, J. Kwieci\'nski, A.D. Martin and P.J. Sutton,
Phys. Rev. {\bf D49} (1994) 4402.

\bibitem{ELR96}
K.J. Eskola, A.V. Leonidov and P.V. Ruuskanen, Nucl. Phys. {\bf B481}
(1996) 704.

\bibitem{GBW_glue}
K. Golec-Biernat and M. W\"usthoff, Phys. Rev. {\bf D60} (1999)
114023-1.

\bibitem{KL01}
D. Kharzeev and E. Levin, Phys. Lett. {\bf B523} (2001) 79.

\bibitem{IN02}
I.P. Ivanov and N.N. Nikolaev, Phys. Rev. {\bf D65} (2002) 054004.


\bibitem{GRV95}
M. Gl\"uck, E. Reya and A. Vogt, Z. Phys. {\bf C67} (1995) 433.

\bibitem{GRV98}
M. Gl\"uck, E. Reya and A. Vogt, Eur. Phys. J. {\bf C5} (1998) 461.


\bibitem{D0_data}
V.M. Abazov et al. (D0 collaboration), Phys. Rev. Lett. {\bf 94} (2005) 
221801.

\end{thebibliography}
\end{document}